\documentclass[11pt,document,nofootinbib,superscriptaddress,onecolumn,preprintnumbers,balancelastpage]{article}
\pdfoutput=1
\usepackage{jheppub}
\usepackage{graphicx}
\usepackage{epstopdf}
\usepackage{dcolumn}
\usepackage{bm}
\usepackage{hyperref}
\usepackage{booktabs}
\usepackage[dvipsnames]{xcolor}
\usepackage{amsmath}
\usepackage{cancel}
\usepackage{xpatch}
\usepackage{mathtools}
\definecolor{c1}{rgb}{0.368417, 0.506779, 0.709798}
\definecolor{c2}{rgb}{0.880722, 0.611041, 0.142051}
\definecolor{c3}{rgb}{0.560181, 0.691569, 0.194885}
\definecolor{c4}{rgb}{0.922526, 0.385626, 0.209179}
\definecolor{c5}{rgb}{0.528488, 0.470624, 0.701351}
\definecolor{c6}{rgb}{0.772079, 0.431554, 0.102387}
\definecolor{c7}{rgb}{0.363898, 0.618501, 0.782349}
\definecolor{turq}{rgb}{0.181,0.638,0.594}
\definecolor{pink}{rgb}{1.000,0.54,0.8}
\definecolor{purple}{RGB}{155,100,155}
\definecolor{gray}{RGB}{128,128,128}
\definecolor{lightBlue}{RGB}{148,179,229}
\definecolor{lightRed}{RGB}{213,157,131}
\definecolor{violet}{RGB}{130,121,173}
\definecolor{gold}{RGB}{255,191,0}

\def\m{\mu}

\def\beq{\begin{align}}
\def\eeq{\end{align}}
\newcommand{\gsim}{ \mathop{}_{\textstyle \sim}^{\textstyle >} }
\newcommand{\lsim}{ \mathop{}_{\textstyle \sim}^{\textstyle <} }
\newcommand{\vev}[1]{ \left\langle {#1} \right\rangle }

\newcommand{\EV}{ {\rm eV} }

\newcommand{\GEV}{ {\rm GeV} }

\def\GeV{{\rm GeV}}

\newcommand{\KH}[1]{{\color{blue} KH:\bf~#1}}

\makeatletter
\g@addto@macro\bfseries{\boldmath}
\makeatother

\title{

Leptogenesis in Parity Solutions to the Strong CP Problem and Standard Model Parameters

}

\author[1,2]{Juanca Carrasco-Martinez}
\author[3]{David I. Dunsky}
\author[1,2]{Lawrence J. Hall}
\author[4,5,6,7]{Keisuke Harigaya}

\affiliation[1]{Department of Physics, University of California, Berkeley, California 94720, USA}
\affiliation[2]{Theoretical Physics Group, Lawrence Berkeley National Laboratory, Berkeley, California 94720, USA}
\affiliation[3]{Center for Cosmology and Particle Physics, Department of Physics,
New York University, New York, NY 10003, USA}
\affiliation[4]{Department of Physics, University of Chicago, Chicago, IL 60637, USA}
\affiliation[5]{Enrico Fermi Institute and Kavli Institute for Cosmological Physics, University of Chicago, Chicago, IL 60637, USA}
\affiliation[6]{Kavli Institute for the Physics and Mathematics of the Universe (WPI),
The University of Tokyo Institutes for Advanced Study,
The University of Tokyo, Kashiwa, Chiba 277-8583, Japan}
\affiliation[7]{Theoretical Physics Department, CERN, Geneva, Switzerland}

\abstract{We study the simplest theories with exact spacetime parity that solve the strong CP problem and successfully generate the cosmological baryon asymmetry via decays of right-handed neutrinos. Lower bounds are derived for the masses of the right-handed neutrinos and for the scale of spontaneous parity breaking, $v_R$. For generic thermal leptogenesis, $v_R \gsim 10^{12}$ GeV, unless the small observed neutrino masses arise from fine-tuning. We compute $v_R$ in terms of the top quark mass, the QCD coupling, and the Higgs boson mass and find this bound is consistent with current data at $1 \sigma$. Future precision measurements of these parameters may provide support for the theory or, if $v_R$ is determined to be below $10^{12}$ GeV, force modifications. However, modified cosmologies do not easily allow reductions in $v_R$ -- no reduction is possible if leptogenesis occurs in the collisions of domain walls formed at parity breaking, and at most a factor 10 reduction is possible with non-thermal leptogenesis. Standard Model parameters that yield low values for $v_R$ can only be accommodated by having a high degree of degeneracy among the right-handed neutrinos involved in leptogenesis. If future precision measurements determine $v_R$ to be above $10^{12}$ GeV, it is likely that higher-dimensional operators of the theory will yield a neutron electric dipole moment accessible to ongoing experiments. This is especially true in a simple UV completion of the neutrino sector, involving gauge singlet fermions, where the bound from successful leptogenesis is strengthened to $v_R \gsim 10^{13}$ GeV.   }

\date{\today}

\begin{document}
\maketitle
\flushbottom

\newpage

\section{Introduction}
Leptogenesis requires neutrino masses to be Majorana and CP violation in the lepton sector.  Observations of neutrinoless double beta decay and different oscillation rates for neutrinos and anti-neutrinos would therefore maintain strong interest in leptogenesis. However, the SM viewed as an effective field theory is also expected to have these properties, so Majorana neutrino masses and CP violation in neutrino oscillations are hardly evidence for leptogenesis. Evidence from neutrino data could emerge in the context of an underlying theory where the number of parameters in the lepton sector is reduced, either by flavor symmetries \cite{Barbieri:1999ma} or by assuming that some entries in the neutrino Yukawa matrix are too small to be relevant \cite{Frampton:2002qc, Harigaya:2012bw}.  Here we try an alternative direction.

A particularly simple and motivated setting for leptogenesis is provided by theories with an underlying parity symmetry. The new fermions required for leptogenesis are right-handed neutrinos, required by the parity transformation $\nu_L \leftrightarrow \nu_R$. Theories with parity are strongly motivated by the elegance of the matter representation \cite{Pati:1974yy,Fritzsch:1974nn} and as a solution to the strong CP problem \cite{Beg:1978mt,Mohapatra:1978fy,Babu:1989rb}. Minimal theories that solve the strong CP problem via an exact parity symmetry double the weak gauge group to $SU(2)_L \times SU(2)_R$ and have a single Higgs doublet for each $SU(2)$, $H_L$ and $H_R$. Parity and $SU(2)_R$ breaking occur spontaneously, with $\vev{H_R} = v_R$ generated radiatively by the Higgs Parity mechanism \cite{Hall:2018let}. The effective theory below $v_R$ is essentially the Standard Model (SM), and $v_R$ can be computed as a function of SM parameters because it is the scale where the SM Higgs quartic coupling vanishes.  There are large uncertainties due to experimental uncertainties in the top quark mass, $m_t$, and the QCD coupling, $\alpha_s$. Currently, $v_R$ lies in the range  $(3 \times 10^{10} - 3 \times 10^{13})$ GeV, at $1 \sigma$.

In these theories Majorana masses for the light neutrinos arise from both the seesaw mechanism by the exchange of $\nu_R$, and from a direct dimension-5 Weinberg operator contribution. The coefficient of the Weinberg operator is determined by $v_R$ and by the size of the right-handed neutrino masses $M_i$, giving
 \begin{align}
 \label{eq:mnudir}
      m_i^{\rm dir} = \frac{v_L^2}{v_R^2} \; M_i. 
\end{align}
Hence the thermal leptogenesis bound $M_1 > 10^9$ GeV from naturalness \cite{Hambye:2003rt}, and the requirement that the direct contribution to the neutrino masses not exceed the observed masses, leads to a naturalness lower bound on $v_R$. If the flavor mixing angles and CP violating phases are of order unity, $v_R$ is predicted to be at the bound. 

In this paper we study leptogenesis in these theories, where radiative spontaneous breaking of exact parity symmetry solves the strong CP problem. We show that successful leptogenesis gives rise to strong bounds on $v_R$ solely from forbidding fine-tunings in the neutrino mass matrix. Such bounds on $v_R$ can be translated to bounds on Standard Model parameters such as $m_t$ and $\alpha_s$ which are associated with $v_R$ as discussed above. 
Moreover, we discuss the implications of these measurements for the estimate of the neutron electric dipole moment that arises from a higher-dimension operator involving $H_R$ and the breaking of parity.

Leptogenesis in the same class of theories was analyzed in~\cite{Dunsky:2020dhn}, coupled with the additional requirement that dark matter is composed of the lightest right-handed neutrino. This dark matter was produced by relativistic freeze-out with subsequent dilution from the decay of a heavier right-handed neutrino or by freeze-in with a low reheating temperature. Here, we remove this dark matter requirement and analyze in detail the lowest possible $v_R$ that allows the right-handed neutrinos to generate the observed baryon asymmetry of the universe through thermal or non-thermal leptogenesis.

In Sec.~\ref{sec:HP} we review the two simplest models of parity restoration that solve the strong CP problem. The Left-Right model has no quarks and charged leptons beyond those of the SM, while in the Mirror model a set of mirror fermions is introduced. In Sec.~\ref{sec:vRHP} we compute $v_R$,
paying particular attention to the dependence on $m_t$ and $\alpha_s$. In Sec.~\ref{sec:numasses} we discuss the origin of neutrino masses; the effective field theory for the neutrino sector is identical in the Left-Right and Mirror models. We derive the relationship between the mass matrix of the active neutrinos, the right-handed neutrino masses and neutrino Yukawa couplings relevant for leptogenesis.  

In Sec.~\ref{sec:vRthermal} we compute the naturalness bound on $v_R$ from thermal leptogenesis in the case that the right-handed neutrinos do not possess a high degree of degeneracy. The bound is very strong, so in Sec.~\ref{sec:vRrelax} we study whether it can be weakened by the decays of a non-thermal population of right-handed neutrinos \ref{subsec:nonthermal}, degeneracy of right-handed neutrinos \ref{subsec:degennuR}, or in the Radiative Singlet Model where the light neutrino masses vanish at tree-level \ref{subsec:SingletModel}. 

Up until this point, we assume parity breaking occurs at a phase transition before inflation, so that the resulting domain walls are inflated away.  In Sec.~\ref{sec:DW} we relax the requirement of exact parity in the Higgs sector so that the spontaneous breaking of parity may occur after inflation, leading to an era with a domain wall network. We investigate whether the bound on $v_R$ could be weakened by having leptogenesis occur during collisions of domain walls.  

A summary of each of these leptogenesis scenarios and their corresponding limits on both the scale of the right-handed neutrino masses, $M_i$, and the scale of right-handed symmetry breaking, $v_R$, is shown in the table below. Conclusions are presented in Sec.~\ref{sec:summary}. 
\begin{table}[ht]
\centering
\begin{tabular}{ |p{6.25cm}||p{3.5cm}|p{3.1cm}|  }
 \hline
 Leptogenesis Scenario  & Lower Limit on $M_i$  &   Lower Limit on $v_R$\\
 \hline
 Thermal Right-Handed Neutrinos, $\nu_R$    &   $\sim 10^9$ GeV (\ref{subsec:M1bound})            &   $\sim 10^{12}$ GeV (\ref{subsec:vRbound}) \\
 Non-thermal  $\nu_R$                       &   $\sim 10^7$ GeV (\ref{subsec:nonthermal})         &   $\sim 10^{11}$ GeV (\ref{subsec:nonthermal}) \\
 Degenerate $\nu_R$                         &   $\sim 2\times 10^3$ GeV (\ref{subsec:degennuR}) &   $\sim 10^{9}$ GeV (\ref{subsec:degennuR}) \\
 Radiative Singlet Model                    &   $\sim 10^{11}$ GeV (\ref{subsec:SingletModel})    &   $\sim 10^{13}$ GeV (\ref{subsec:SingletModel}) \\
 Domain Wall Collisions                          &   $\sim 10^{9}$ GeV  (\ref{sec:DW})                 &   $\sim 10^{12}$ GeV (\ref{sec:DW}) \\
 \hline
\end{tabular}
\label{tab:summary}
\end{table}

\section{Parity Restoration with Minimal Higgs Doublets (Higgs Parity)}
\label{sec:HP}

A remarkable feature of the Standard Model (SM) is that spacetime parity is broken. The central role played by parity conservation in atomic and nuclear physics, and the apparently fundamental nature of the transformation $\bf{r} \rightarrow - \bf{r}$, resulted in parity violation emerging in the mid-1950s as a profound surprise \cite{Lee:1956qn,Okun:2006eb}.  Left-handed fermions appear as doublets of the weak interaction gauge group $SU(2)_L$, while the right-handed fermions do not feel this weak force. For example, $\ell_L = (\nu_L, e_L)$ represent the $SU(2)_L$ doublets of left-handed neutrinos and charged leptons, while $e_R$ are $SU(2)_L$ singlets and right-handed neutrinos have not been discovered.

Although $\nu_L$ and $e_L$ are treated symmetrically by the weak interaction, it is hard to conceive of two particles less like each other than the neutrino and electron.  The differences emerge from the breaking of the weak interaction symmetry by the condensate of the Higgs field $H_L$, which is also a doublet of $SU(2)_L$ and was discovered a decade ago. The potential for the Higgs field is extremely simple involving two real parameters
\begin{align}
    \label{eq:SMHiggspot}
    V_{\rm SM}(H_L) = - \mu^2 \, H_L^\dagger H_L + \frac{\lambda}{2} \,(H_L^\dagger H_L)^2 .
\end{align}

The minimal extension of the Standard Model that allows an excess of baryons over anti-baryons to develop in the early universe is obtained by adding Majorana fermions, $N$, that do not feel the strong or electroweak interactions and have Lagrangian terms
\begin{align}
    \label{eq:NLag}
    L_N =  y \; \ell_L N H_L + \frac{1}{2} M_N \, NN,
\end{align}
where generation indices are omitted. If the couplings $y$ are large enough, $N$ were
produced
during an early era of the hot big bang. As the temperature dropped below their mass, they generate a lepton asymmetry by decays
\begin{align}
    \label{eq:Ndecay}
    N \rightarrow \ell_L H_L, \;\; \overline{\ell_L} \; \overline{H_L},
\end{align}
with different decay rates to the particle and anti-particle modes. The lepton asymmetry is processed to a baryon asymmetry by sphalerons of the $SU(2)_L$ weak interactions.

Parity restoration provides a simple framework for realizing this cosmological scheme of {\it leptogenesis}, as the fermions $N$ are interpreted as right-handed neutrinos, which parity requires.  To construct a theory which is parity invariant it is necessary that, in addition to reflecting the spatial coordinate, the weak interaction of $SU(2)_L$ is transformed into another interaction, $SU(2)_R$, that acts on right-handed fermions. Thus the electroweak gauge group must contain  $SU(2)_L \times SU(2)_R$, and nature must contain $W_R$ gauge bosons of $SU(2)_R$, in addition to the observed $W_L$ gauge bosons of $SU(2)_L$. Under parity, in addition to $\bf{r} \rightarrow - \bf{r}$,
\begin{align}
    \label{eq:Ptrans}
\begin{aligned}
  SU(2)_L \;\; &\leftrightarrow \;\; SU(2)_R, \hspace{0.2in}
  \\
 \nonumber
  \ell_L (2,1) = 
  \begin{pmatrix}
\nu_L\\
e_L
\end{pmatrix} \;\;\;
  &\leftrightarrow \;\;\; \ell_R (1,2) =
  \begin{pmatrix}
\nu_R \\
\psi_R
\end{pmatrix}.
\end{aligned}
\end{align}

There are two theories of parity restoration that are consistent with thermal leptogenesis. In both theories, the gauge group is $SU(3)_c \times SU(2)_L \times SU(2)_R \times U(1)$ or $SU(4)\times SU(2)_L\times SU(2)_R$. In the minimal version (Model A in~\cite{Hall:2018let}) which we call the left-right (LR) theory,
the Parity partners $(\bar{q},\bar{\ell})$ of the $SU(2)_L$ doublet SM fermions $(q,\ell)$ are $SU(2)_R$ doublets and are right-handed SM fermions except for $\nu_R$, and the $U(1)$ generator is $B-L$ when acting on SM fermions. In particular, $\psi_R$ are identified as the right-handed charged leptons, $e_R$.
In the other, non-minimal, theory (Model D in~\cite{Hall:2018let}) which we call the Mirror theory in this paper, all of the Parity partners $(q',\ell',\bar{u}',\bar{d}',\bar{e}')$ of the SM fermions $(q,\ell,\bar{u},\bar{d},\bar{e})$ are new particles. These mirror fermions have the same $SU(3)_c\times U(1)_Y$ charges as the SM counterparts, and $(q',\ell')$ are $SU(2)_R$ doublets.
In both theories, the neutrinos $\nu_L$ and $\nu_R$ have the same mass terms and the same interactions relevant for leptogenesis, so the analysis and results of this paper apply to both the LR theory and the Mirror theory.

There are three other closely related theories of parity restoration. Two (Models B and C in~\cite{Hall:2018let}) have a gauge group $SU(3)_c \times SU(2)_L \times SU(2)_R \times U(1)$, but differ from the LR and mirror theory in whether parity conjugates the color and hypercharge quantum numbers. Another one, with electroweak symmetry completely mirrored, has a gauge group $SU(3)_c \times SU(2)_L \times SU(2)_R \times U(1)_Y \times U(1)_{Y'}$, fermions are doubled, and $\psi_R$ are mirror leptons charged under Mirror QED, $e'_R$~\cite{Barr:1991qx}.
However, in these three theories, the lightest extra quark is stable and necessarily produced at the high temperatures necessary for leptogenesis. At the QCD phase transition, these quarks hadronize with SM quarks, forming heavy fractionally charged particles whose terrestrial abundance is highly constrained \cite{Dunsky:2018mqs}. Consequently, the cosmology of these theories
is challenging to reconcile with leptogenesis. For this reason, we hereafter mainly focus on the LR and Mirror theories which do not have these problems. (See, however, discussion at the end of Sec.~\ref{subsec:degennuR}.)
The $U(1)\times U(1)'$ symmetry of the Mirror electroweak theory may be spontaneously broken down to $U(1)$ to obtain the LR or Mirror theory.

\begin{figure}[tb]
    \centering
    \includegraphics[width=1.0\textwidth]{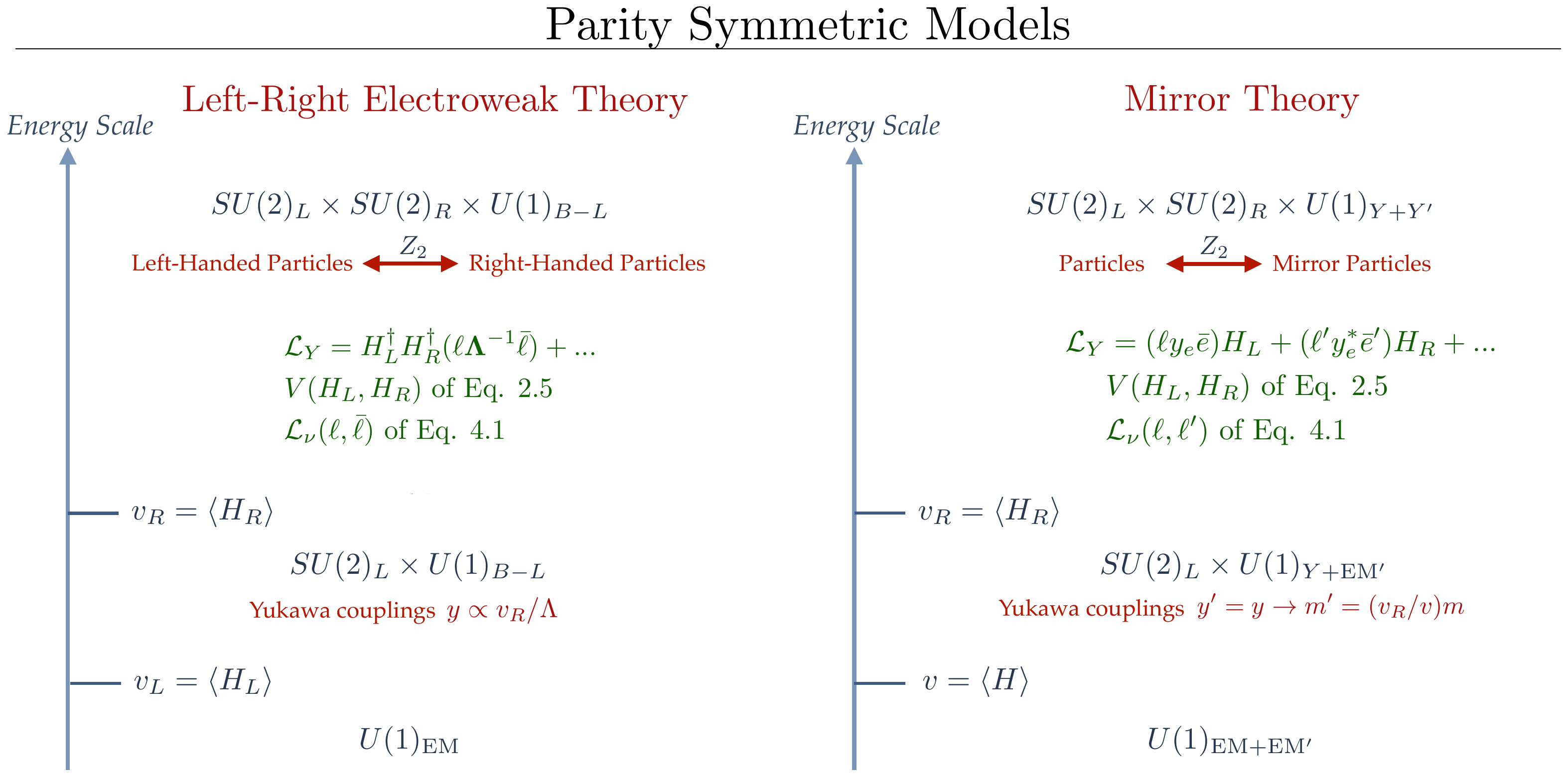} 
    \caption{
    Overview of the two parity-restoring models with minimal Higgs sectors. The left-panel shows the `Left-Right' theory, which has gauge symmetry $SU(3)_c \times SU(2)_L \times SU(2)_R \times U(1)_{B-L}$ at high energies. 
    Parity interchanges left-handed and right-handed quarks and leptons, and is spontaneously broken at $v_R$ by the $H_R$ vev, which also breaks the gauge group: $SU(2)_R \times U(1)_{B-L} \rightarrow U(1)_Y$. SM Yukawa couplings arise from the dimension-five operators shown in green. 
    The right-panel shows the `Mirror' theory which contains mirror quarks and leptons. The strong interaction gauge group may be $SU(3)_c$ or $SU(3) \times SU(3)' \rightarrow SU(3)_c$. Spacetime parity maps SM fields to their chirality-flipped mirror counterparts. Parity is broken at $v_R$ by the vev of $H_R$, with charged mirror fermions acquiring masses $v_R/v$ larger than their SM counterparts, as shown in red.}
    \label{fig:LRvsMirror_Figure}
\end{figure}

Experimental searches for $W_R$ have placed a lower limit on $v_R$ of 13 TeV in the LR theory~\cite{ATLAS:2019lsy}, where the parity partners of the left-handed SM fermions are right-handed SM fermions. In the Mirror theory, the mirror quarks are triplets under QCD, at least at LHC energies, so that the $u'$ mass must be above 2 TeV, forcing $v_R$ above $10^8$ GeV. $SU(2)_R$ must be broken at a larger scale than $SU(2)_L$. A key question 
is how the Higgs fields transform under the gauge group 
$SU(2)_L \times SU(2)_R$.  In this paper we explore the possibility that the Higgs system is minimal, namely under parity
\begin{align}
  H_L(2,1) \;\; \leftrightarrow \;\; H_R^\dagger(1,2),
\end{align}
leading to a doubling of the SM Higgs potential and the addition of a quartic coupling
\begin{align}
    \label{eq:SMHiggspot}
    V(H_L,H_R) = \Biggl( - \mu^2 \, H_L^\dagger H_L + \frac{\lambda}{2} \,(H_L^\dagger H_L)^2 \Biggr) + \Biggl(
    - \mu^2 \, H_R^\dagger H_R + \frac{\lambda}{2} \,(H_R^\dagger H_R)^2\Biggr) + \xi \, H_L^\dagger H_L H_R^\dagger H_R.
\end{align}
At tree-level there is no realistic vacuum solution of this potential; either $H_L$ and $H_R$ have equal vacuum values, $v_R = v_L$, or one of them vanishes, $v_L=0$ \cite{Barbieri:2005ri}.  However, radiative corrections allow a realistic vacuum with parity spontaneously broken at the scale where the SM quartic coupling passes through zero, giving a $1 \sigma$ range of
\begin{align}
    \label{eq:vRinHP}
  v_R \simeq (3 \times 10^{10} - 3 \times 10^{13}) \; \GeV,  
\end{align}
as discussed in the next section. This theory of Higgs Parity \cite{Hall:2018let} is the minimal Higgs sector with parity restoration and, unlike conventional LR theories with $SU(2)_L \times SU(2)_R$, the parameters of the Higgs potential are real. Parity solves the strong CP problem in the LR theory, with radiative corrections to $\bar{\theta}$ arising at 2-loops \cite{Hall:2018let, Hisano:2023izx} offering the possibility of detection of a neutron electric dipole moment in upcoming experiments \cite{Alarcon:2022ero}.\footnote{A non-minimal alternative is to introduce soft breaking of parity in the scalar potential, $\delta V = \m^2 (H_L^\dagger H_L - H_R^\dagger H_R)$. Realistic vacua then result at tree-level, with any value of $v_R$ above the experimental bound. Indeed, this was the first model constructed with parity solving the strong CP problem \cite{Babu:1989rb}. This soft breaking in the electroweak Higgs potential could arise from spontaneous breaking in some other sector of the theory.   In this paper we restrict our attention to the minimal electroweak Higgs potential of (\ref{eq:SMHiggspot}) with parity exact.}
Strong CP is also solved in the Mirror theory, providing QCD is not mirrored \cite{Barr:1991qx,Dunsky:2019api} or arises from the breaking of $SU(3) \times SU(3)'$ to the diagonal sum \cite{Barr:1991qx, Bonnefoy:2023afx}. In these cases the radiative corrections to $\bar{\theta}$ are negligible, as in the SM.

The theory with Higgs Parity, whether LR or mirror, is an EFT and is expected to have higher-dimension operators that generate a non-zero value of $\bar{\theta}$ once parity is spontaneously broken. These first arise at dimension 6 \cite{Hall:2018let}
\begin{align}
    \label{eq:dim6}
  {\cal L}_6 = \frac{C}{M_{Pl}^2} (H_L^\dagger H_L - H_R^\dagger H_R) \; G \tilde{G} \; + \; ... 
\end{align}
where $G$ is the QCD field strength. There are also relevant operators that correct the quark Yukawa couplings at order $v_R^2/M_{Pl}^2$. The resulting bound on $v_R$ is highly uncertain, as the UV completion at the Planck scale $M_{Pl}$ is not known. For example, with $C=1$ the bound is $v_R \lesssim (10^{12} - 10^{13})$ GeV for $M_{Pl}$ in the range of the Planck or reduced Planck mass. Thus, as $v_R$ is increased above about $(10^{13} - 10^{14})$ GeV, a ``Little strong CP problem" emerges. However, it may be that the LR or mirror theory has a UV completion before the Planck scale that leads to $C\ll 1$, allowing these larger values of $v_R$.  For example, if parity is embedded into $SO(10) \times CP$ \cite{Hall:2018let} we expect $ C \sim v_{10}/M_{Pl}$, where $v_{10}$ is the $SO(10)$ breaking scale. Similarly $C\ll 1$ if supersymmetry is encountered between $v_R$ and $M_{Pl}$.  In the coming decade, experiments searching  for the neutron electric dipole moment will improve sensitivity by a factor of 30 \cite{Alarcon:2022ero}, thereby probing values of $v_R$ higher by a factor of 5. Thus, there is a good chance of discovery if $v_R$ is in the range of $(10^{12} - 10^{14})$ GeV and the LR or mirror model is UV completed at the Planck scale, and there is also a chance for discovery if $v_R$ is larger and a UV completion with $C \ll 1$ comes earlier.

When gauge singlets $N$ are added to the SM, and thermal leptogenesis proceeds via the interactions of (\ref{eq:NLag}), there is a bound on the mass of the lightest $N$ involved in the process, $M_N \gsim 10^9$ GeV as discussed in Sec.~\ref{subsec:M1bound}.  This can be relaxed in theories where the decaying $N$ have a non-thermal abundance or where there is degeneracy among the $N$. In this paper we study bounds on $m_{\nu_R}$ and $v_R$ from leptogenesis in parity symmetric theories with minimal Higgs fields. In addition to the seesaw contribution to the light neutrino mass, which is inversely proportional to $m_{\nu_R}$, there is a direct term from a dimension-5 operator, $(v_L^2/v_R^2)m_{\nu_R}$, proportional to $m_{\nu_R}$ as discussed in Sec.~\ref{sec:numasses}.  Forbidding a fine-tuned cancellation between these contributions, we derive a lower bound of $v_R > 3 \times 10^{11}$ GeV, in the middle of the range (\ref{eq:vRinHP}) allowed by the Higgs Parity theory.

\section{The Scale of $SU(2)_R$ in Higgs Parity}
\label{sec:vRHP}
We begin with a brief review of how the scale $v_R$ is correlated with the values of the top quark mass, $m_t$, and the QCD coupling, $\alpha_s$ in the Higgs Parity theory~\cite{Hall:2018let}. The minimal Higgs potential, (\ref{eq:SMHiggspot}), is usefully rewritten as 
\begin{align}
\label{eq:potential}
V(H_L,H_R) = - \mu^2 \left( H_L^\dagger H_L + H_R^\dagger H_R \right) + \frac{\lambda}{2} \left( H_L^\dagger H_L + H_R^\dagger H_R \right) ^2 + \lambda' \; H_L^\dagger H_L  H_R^\dagger H_R.
\end{align}
We assume that the mass scale $\mu$ is much larger than the electroweak scale, $v_L$.

With positive $\mu^2$, $H_R$ obtains a large vacuum expectation value $\vev{H_R} = \mu  / \lambda^{1/2} \equiv v_R$ and spacetime parity is spontaneously broken. After integrating out $H_R$ at tree-level, the low energy effective potential of $H_L$ is 
\begin{align}
V_{\rm LE}(H_L) = \lambda' \; v_R^2  \;  H_L^\dagger H_L - \lambda' \left(1  + \frac{\lambda'}{2 \lambda} \right) (H_L^\dagger H_L)^2 .
\end{align}
The hierarchy $v_L \ll v_R$ is obtained only if the quadratic term is small, which requires a small value of $\lambda' \sim - v_L^2/v_R^2$. The quartic coupling of the Higgs $H_L$, $\lambda_{\rm SM}$, is then very small at the symmetry breaking scale $v_R$.
The nearly vanishing quartic coupling can be understood by an approximate global $SU(4)$ symmetry under which $(H_L,H_R)$ forms a fundamental representation. For $|\lambda'| \ll 1$ the potential in Eq.~(\ref{eq:potential}) becomes $SU(4)$ symmetric. The $SU(4)$ symmetry is spontaneously broken by $\vev{H_R}$ and the SM Higgs is understood as a Nambu-Goldstone boson with vanishing potential.

At tree-level the potential still leads to $\vev{H_L} = \vev{H_R} = v_R$ because of the small quartic coupling. However, for extremely small $\lambda'$, vacuum alignment in the $SU(4)$ space is fixed by quantum corrections which violate the $SU(4)$ symmetry. The dominant effect is renormalization group running from energy scale $v_R$ down to $v_L$. The top quark contribution dominates over the gauge contribution and generates a positive quartic coupling $\lambda_{\rm SM}(v_L) \simeq 0.1$, and creates the minimum of the potential at $v_L \ll v_R$. From the perspective of running from low to high energy scales, the scale at which the SM Higgs quartic coupling nearly vanishes is the scale $v_R$. Threshold corrections to $\lambda_{\rm SM}(v_R)$ are computed in~\cite{Dunsky:2019api,Hall:2019qwx} and are typically $O(10^{-3})$.

The vacuum alignment can be also understood in the following way.
For $\lambda' > 0$, the minima of the potential are $\left( \vev{H_L}, \vev{H_R} \right)= \left( v_R,0 \right)$ and $\left( 0,v_R \right)$, where $v_R\equiv \mu / \lambda^{1/2}$, and the mass of Higgses are of order $\mu$. For $\lambda' < 0$, the minima are $\vev{H_L} = \vev{H_R} \sim v_R$. None of these minima for $\lambda>0$ and $\lambda'<0$ give rise to an $SU(2)$ sector broken at a scale much lower than the other.
To obtain a viable vacuum, we need $\lambda' \simeq 0$, for which the potential has an accidental $SU(4)$ symmetry and nearly degenerate vacua with $v_L^2 = v_R^2 = \mu^2/\lambda$. In this case, quantum corrections must be taken into account to determine the orientation of the vacuum. The dominant effect is given by the top quark Yukawa coupling, which leads to a Colemann-Weinberg potential that in the limit that $\lambda' = 0$ orients the vev entirely in the $H_R$ or $H_L$ direction.  However, a small negative $\lambda'$ slightly destabilizes the vacuum with all the vev in $H_R$ to give $\left( \vev{H_L}, \vev{H_R} \right)= \left( v_L ,v_R \right)$ with $v_L \ll v_R$. There is also a physically equivalent vacuum with the L and R labels interchanged: we define L by $v_L \ll v_R$.   

\begin{figure}[tb]
    \centering
    \begin{minipage}{0.5\textwidth}
        \centering
        \includegraphics[width=.95\textwidth]{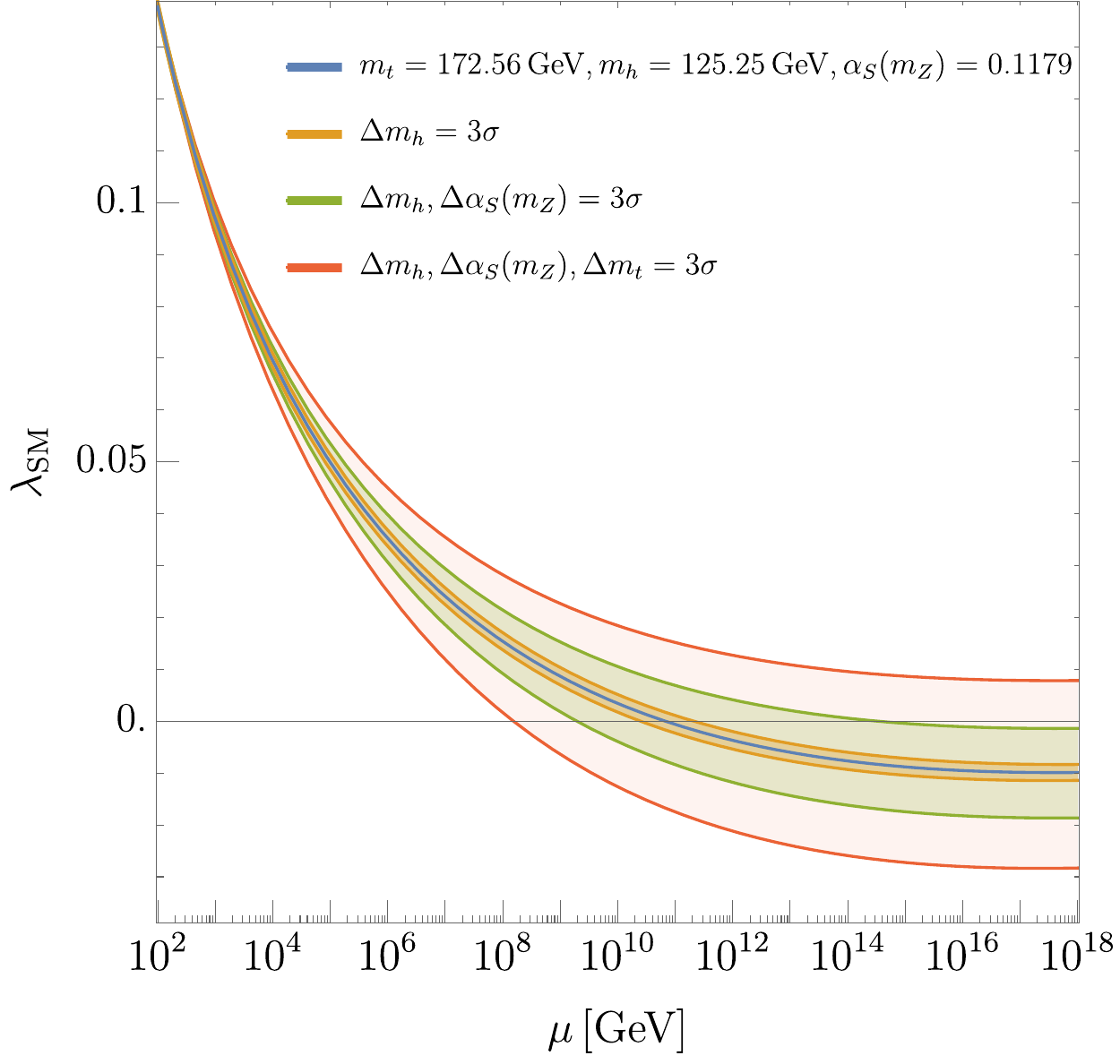} 
    \end{minipage}\hfill
    \begin{minipage}{0.5\textwidth}
        \centering
        \includegraphics[width=.95\textwidth]{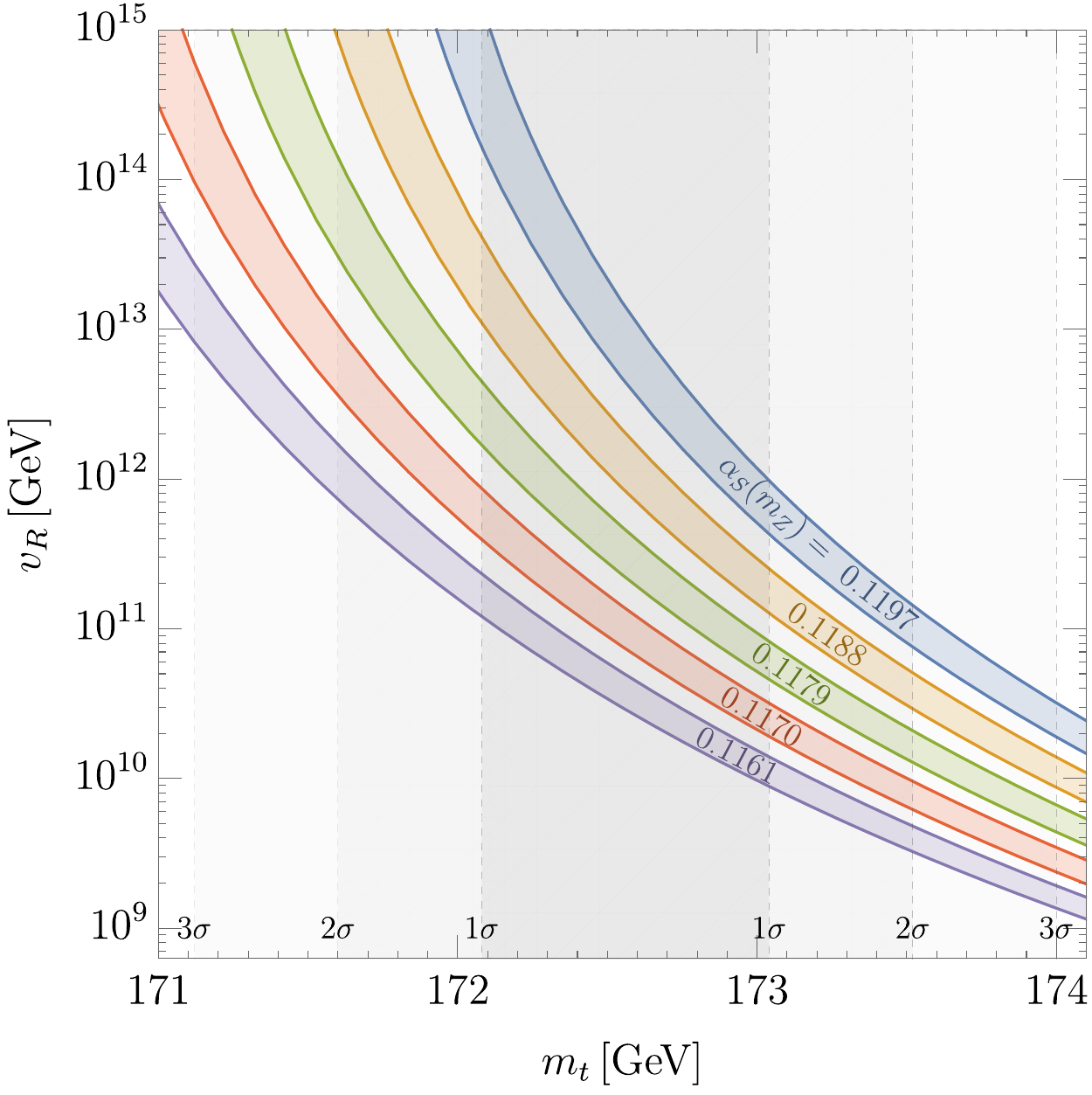} 
    \end{minipage}
    \caption{\small (\textbf{Left}) Running of the SM quartic coupling. (\textbf{Right}) Predictions from the Higgs Parity mechanism for the scale $v_R$ as a function of the top quark mass, $m_t$. Contours of $\alpha_S(M_Z)$ show how the prediction changes with the uncertainty in the QCD coupling constant. The thickness of each contour corresponds to the current $1\sigma$ uncertainty in $m_h$ of $\pm 170$ MeV. Both panels are for the LR model.}
    \label{fig:vpPrediction}
\end{figure}

Between the electroweak scale and the scale of parity restoration, $v_R$, the running of the Higgs quartic coupling $\lambda_{\rm SM}$ is exactly the same as in the SM. We follow the computation in~\cite{Buttazzo:2013uya} and show the running in the left panel of Fig.~\ref{fig:vpPrediction} for a range of values for the top quark mass $m_t = (172.56\pm 0.48)$ GeV, QCD coupling constant at the $Z$ boson mass $\alpha_S(m_Z) = (0.1179 \pm 0.0009)$, and Higgs mass $m_h = (125.25 \pm 0.17)$ GeV. 

The value of the SM quartic coupling at the scale $v_R$ is not exactly zero because of the threshold correction~\cite{Dunsky:2019api},
\begin{align}
\lambda_{\rm SM}(v_R) \simeq - \frac{3}{8\pi^2} y_t^4 \, {\rm ln} \frac{e}{y_t} + \frac{3}{128\pi^2} (g^2 + {g'}^2)^2 \left( {\rm ln} \frac{e\sqrt{2}}{\sqrt{g^2 + {g'}^2 }} -  {\rm ln} \frac{g^2}{\sqrt{g^4 - g'^4}} \right) + \frac{3}{64\pi^2} g^4 \, {\rm ln} \frac{e\sqrt{2}}{ g},
\label{eq:thresholdCorrectionvR}
\end{align}
where the $\overline{\rm MS}$ scheme is assumed.%
\footnote{Here it is assumed that the top Yukawa is simply given by $y_t H_L q \bar{u} + y_t H_R \bar{q} u$. In the LR theory, it is possible that $u$ and $\bar{u}$ has a Dirac mass term. The threshold correction for this case is different from Eq.~\eqref{eq:thresholdCorrectionvR}, but this only makes the prediction on $v_R$ for given SM parameters smaller~\cite{Hall:2019qwx}. In Model D, Yukawa interactions of $H_L q u + H_R \bar{q} \bar{u}$ are allowed by the gauge symmetry. As is shown in Appendix~\ref{sec:model dependence}, the prediction on $v_R$ for given SM parameters can become only smaller also in this case.}
The prediction for the scale $v_R$ is shown in the right panel of Fig.~\ref{fig:vpPrediction} as a function of $m_t$. Colored contours show how the prediction for $v_R$ changes when the QCD coupling constant varies by $\pm 1\sigma$ and $\pm 2\sigma$ deviations about its mean. The thickness of each curve corresponds to the $1\sigma$ uncertainty in the measured Higgs mass, $m_h = (125.25 \pm 0.17)$ GeV. With 2$\sigma$ uncertainties, $v_R$ can be as low as $3 \times 10^9$ GeV. Future measurements of SM parameters can pin down the scale $v_R$ with an accuracy of a few tens of percent~\cite{Dunsky:2019api}.

\begin{figure}[tb]
    \centering
    \begin{minipage}{0.5\textwidth}
        \centering
        \includegraphics[width=.95\textwidth]{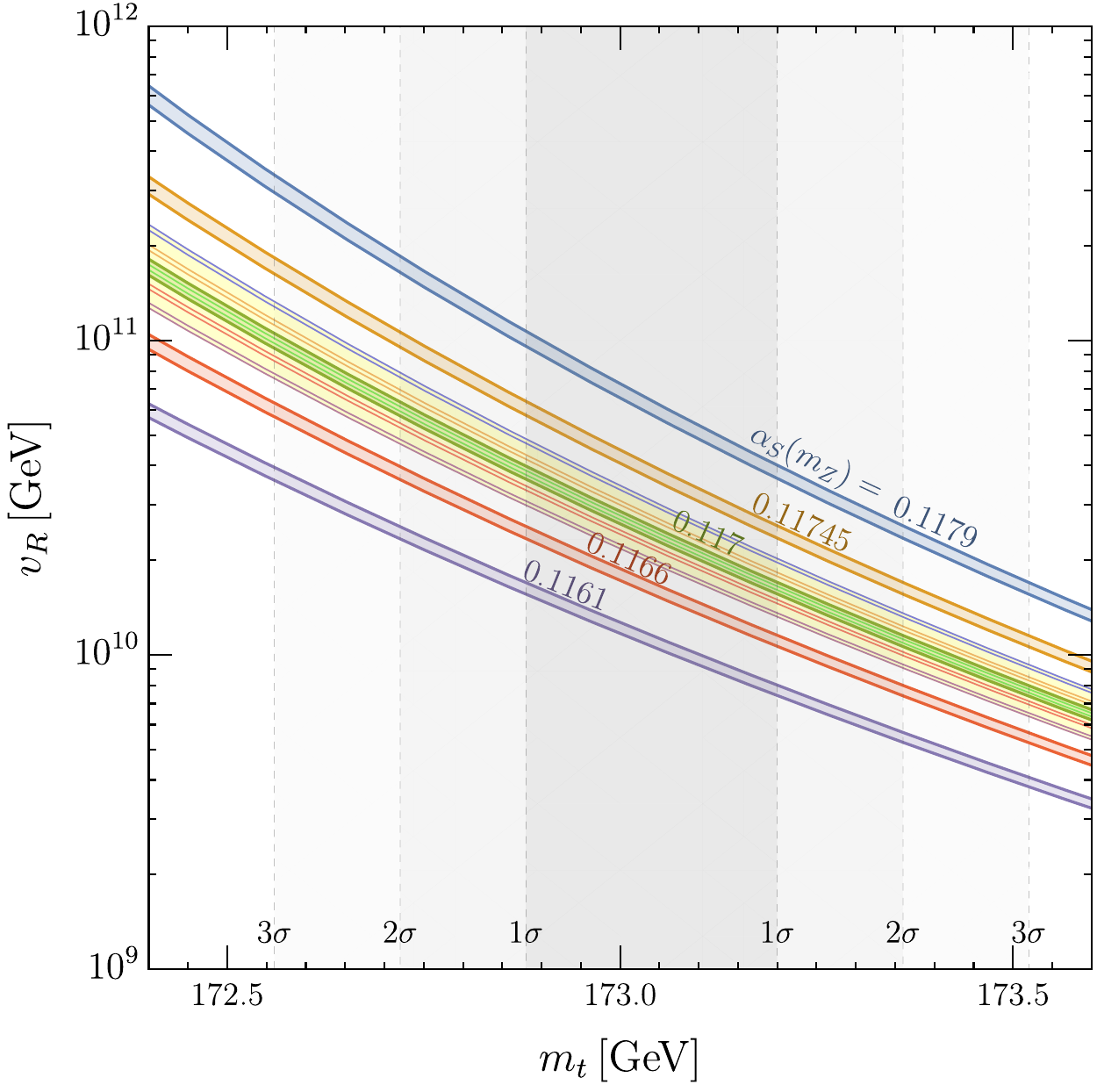} 
    \end{minipage}\hfill
    \begin{minipage}{0.5\textwidth}
        \centering
        \includegraphics[width=.95\textwidth]{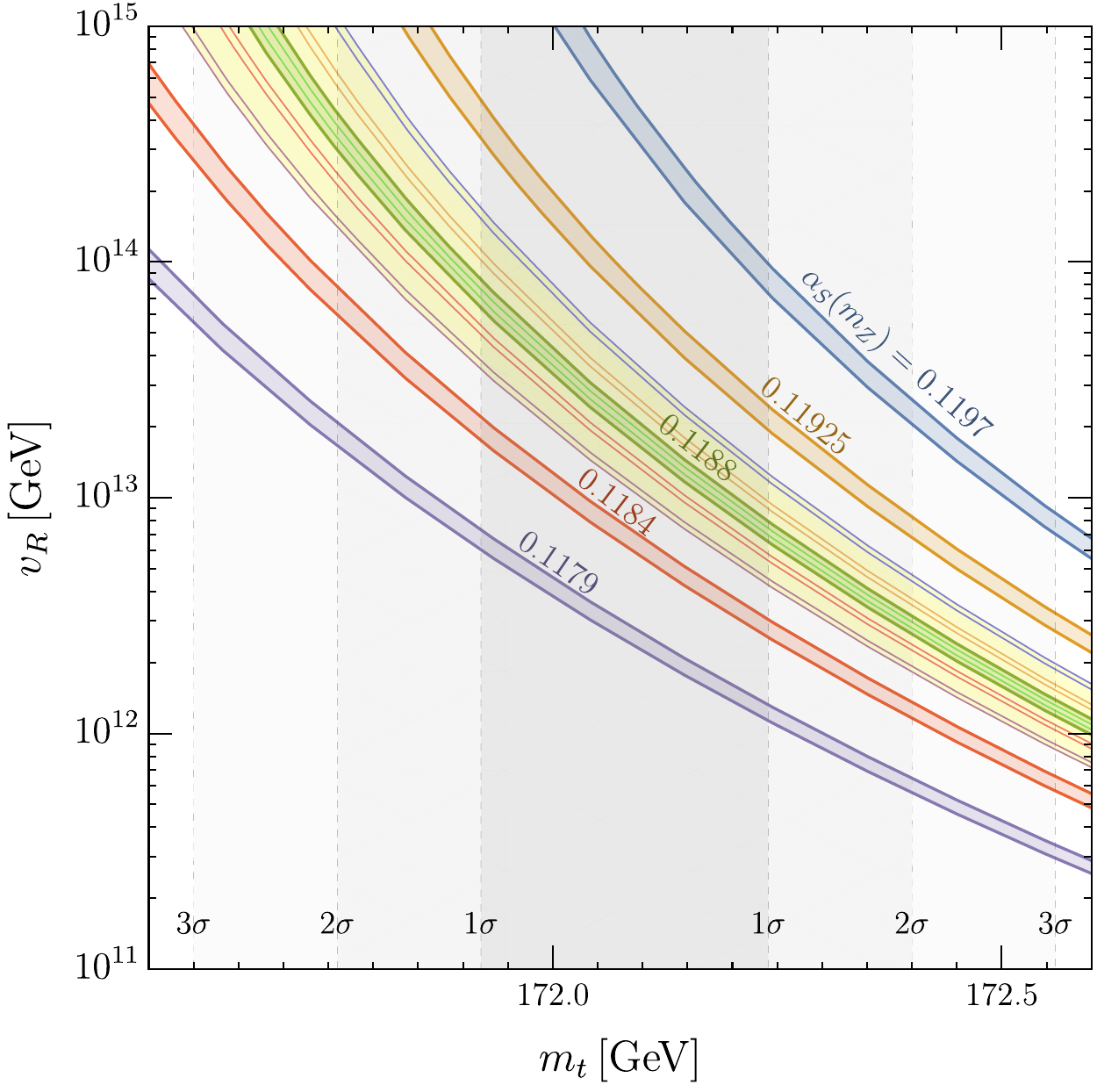} 
    \end{minipage}
    \caption{Future determinations of $v_R$ in the LR model from improved measurements of SM parameters. The figures are analogous to the right panel of Fig.~\ref{fig:vpPrediction} but with improved uncertainties. The labelled colored bands and the uncertainties on the $m_t$ axis are for expected improvements over the next decade, with uncertainties of $\pm 160$ MeV for $m_t$, $\pm 0.00045$ for $\alpha_S(M_Z)$, and $\pm 30$ MeV for $m_h$. The yellow band highlights the even finer precision expected with a next generation $e^+ e^-$ collider. The 3$\sigma$ uncertainty range of $m_t$ from a future $e^+ e^-$ collider is roughly the same as the $1\sigma$ uncertainty range from the High Luminosity LHC, shown as a vertical dark-gray band. \textbf{Left}: Centered on $m_t (\alpha_s)$ that are $1 \sigma$ above (below) the current values.   \textbf{Right}: Centered on $m_t (\alpha_s)$ that are $1 \sigma$ below (above) the current values. }
    \label{fig:vpPredictionFuture}
\end{figure}

How accurately will $v_R$ be determined by future measurements of $m_t, \alpha_s$ and $m_h$? At the present central values, varying both $m_t$ and $\alpha_s$ by the current $1 \sigma$ uncertainties of 480 MeV and 0.0009 gives a range in $v_R$ of about three orders of magnitude, from about $3 \times 10^{10}$ GeV to $3 \times 10^{13}$ GeV.  Studies of expected reductions in these uncertainties \cite{Schwienhorst:2022yqu, dEnterria:2022hzv, Begel:2022kwp, Dawson:2022zbb} offer the possibility of a large improvement in the determination of $v_R$.  Over the coming decade, improvements by a factor 3 on $\delta m_t$ and 5 on $m_h$ from High Luminosity LHC, and a factor 2 on $\delta \alpha_s$, will shrink  the $1 \sigma$ range for $v_R$ to about one order of magnitude.  
However, the future allowed range for $v_R$ depends strongly on the results for the central values of $m_t$ and $\alpha_s$. From the right panel of Fig.~\ref{fig:vpPrediction}, it is apparent that the determination of $v_R$ becomes more precise at higher values of $m_t$, where high values of $v_R$ can be reliably excluded. In the left panel of Fig.~\ref{fig:vpPredictionFuture}, we assume central values for $m_t (\alpha_s)$ that are $1 \sigma$ above (below) the current values and find that the full $1\sigma$ range for $v_R$ is only about a factor of 3. In the right panel, we assume central values for $m_t(\alpha_s)$ that are $1\sigma$ below (above) the current values and find that the full $1 \sigma$ range for $v_R$ is still 2 orders of magnitude. 

At a future $e^+e^-$ collider, a top threshold scan could give a further factor of three reduction in $\delta m_t$, high statistics $Z$ measurements  could reduce $\delta \alpha_s$ by a further factor of 4, and the uncertainty on $m_h$ would be around 10 MeV. The yellow bands of Fig.~\ref{fig:vpPredictionFuture} indicate the uncertainty in $v_R$ with next generation colliders. With present central values for these parameters, the full $1 \sigma$ range for $v_R$ would then be about a factor of two, whereas the ranges for other choices of the central values can be seen in the right panel of Fig.~\ref{fig:vpPredictionFuture}.

 Using an analytic approximation, we find uncertainties in $\log_{10} v_R$ of
\begin{align}
\label{eq:vRuncert}
\delta (\log_{10} v_R) \; = \; (\log_{10} v_R - 7.9) \left( - 0.23 \; \frac{\delta m_t}{0.48 \GeV}, \;\; 0.15 \; \frac{\delta_{\alpha_s}}{0.0009}  \right)
\end{align}
from uncertainties $(\delta m_t, \delta_{\alpha_s})$. This is a good approximation for $10^9 \, \GeV < v_R < 10^{14} \, \GeV$, and the first factor shows how the uncertainty grows with $v_R$. Using this approximation, at the current central values the $1\sigma$ uncertainties in $\log_{10} v_R$ are $(\pm 0.83, \pm 0.54)$ now, $(\pm 0.28, \pm 0.18)$ after High Luminosity LHC, and $(\pm 0.09, \pm 0.07)$ after runs at a future $e^+e^-$ collider.  The latter case corresponds to $\delta v_R/v_R = (\pm 0.21, \pm 0.16)$. 

\section{Neutrino Masses with Higgs Parity}
\label{sec:numasses}

In both the LR and Mirror theories with Higgs Parity, neutrino masses arise from operators of dimension-5
\begin{align}
{\cal L}_\nu &= -\frac{1}{2M} \left( \ell_i c^*_{ij} \ell_j \; H_L H_L  +   \bar{\ell}_i c_{ij}\bar{\ell}_j \; H_R H_R \right)  +  \frac{1}{M} \,  \ell_i b_{ij} \bar{\ell}_j  \; H_L H_R + {\rm h.c.},
\label{eq:yukawaNu}
\end{align}
where $M$ is real, $c_{ij}$ is symmetric and $b_{ij}$ is Hermitian. The leptons, which transform under $SU(2)_L \times SU(2)_R$ as $\ell(2,1)$ and $\bar{\ell}(1,2)$, are described by two-component left-handed Weyl fields, so that the parity transformation is $\ell \leftrightarrow \bar{\ell}^\dagger$, and $H_L \leftrightarrow H_R^\dag$.
Previously we called the lepton doublets of the Mirror theory $\ell'$; now we call them $\bar{\ell}$, so that our analysis applies to both LR and Mirror theories. In both theories, we write the neutral field in this doublet as $\bar{\nu}$, and describe the corresponding state as a right handed neutrino, $\nu_R$.

The operators of (\ref{eq:yukawaNu}) can arise from exchanges of heavy fermions, transforming under $SU(2)_L \times SU(2)_R$ as $(1,1)$ or $(3,1) + (1,3)$ for the lepton number violating case and $(1,1)$ or $(2,2)$ for the lepton number conserving operator~\cite{Hall:2018let,Hall:2019qwx}. In section \ref{subsec:SingletModel} we explore interesting features of the theory where both types of operator are generated by three Majorana singlets $S(1,1)$. The operators can also arise from the exchange of massive scalars transforming as $(3,1) + (1,3)$  ($(2,2)$) for the lepton number violating (conserving) operators. If the masses of the heavy exchanged particles are larger than $v_R$, then the effective theory (\ref{eq:yukawaNu}) applies at scales of $v_R$ and below. Even if these masses are below $v_R$, providing they are the largest mass terms in the neutral fermion mass matrix, the effective theory relevant for neutrino masses and leptogenesis is given by (\ref{eq:yukawaNu}) with $H_R$ replaced by $v_R$.

The effective Lagrangian of \eqref{eq:yukawaNu} leads to a $6 \times 6$ neutrino mass matrix,
\begin{align}
\begin{array}{c} \big( \begin{array}{cc}\nu  _i   & \bar{\nu}_i \end{array} \big) \\ {} \end{array}
  \begin{pmatrix}
 M_{ij} \,v_L ^2 /v_R ^2 \hspace{0.25in} \, & y_{ij} v_L \\
 y_{ji} v_L & M^*_{ij}
\end{pmatrix} \bigg( \begin{array}{c} 
 \nu _j \\  
 \bar{\nu}_j 
\end{array} \bigg)
  \,,
\end{align}
where $M_{ij} = c_{ij} v_R^2/M$ and $y_{ij} = b_{ij} v_R/M$.  Without loss of generality, we can work in a basis where $c_{ij}$ is diagonal such that
\begin{align}
	M_{ij} &= M_i \, \delta_{ij},
 \label{eq:Mi}
\end{align}
with all $M_i$ real and positive and no summation over indices.  On integrating out the three heavy states assuming $c v_R^2 \gg b v_L v_R$, we obtain a mass matrix for the three light neutrinos:
\begin{align}
	m_{ij} \, &= \, \delta_{ij} \frac{v_L^2}{v_R^2} M_i - y_{ik} \; \frac{1}{M_k} \; y^T_{kj} v_L^2 \,\equiv \, \delta_{ij} \, m_{i}^{\rm dir} - m_{ij}^{\rm ss}.
	\label{eq:numassmatrix}
\end{align}
We call the first term the ``direct" contribution and the second the ``seesaw" contribution.

It is useful to study the direct and seesaw contributions to the light neutrino mass matrix from each $\nu_{R_i}$.  Each $\bar{\nu}_i$ field couples to a single combination of $\ell_j$, which we call $\tilde{\ell}_i$, so that the Yukawa coupling of $\bar{\nu}_i$ can be written as
 \begin{align}
 \label{eq:Lyi}
     {\cal L}_{y_i} = y_i \; \tilde{\ell_i} \,\bar{\nu}_i \,  H + {\rm h.c.}, \qquad \tilde{\ell_i} =  \ell_j \, y_{ji}/y_i, \qquad y_i^2 \equiv \sum_j |y_{ji}|^2,
\end{align}
with $y_i$ real. Thus, each $\nu_{R_i}$ gives mass contributions to two different states, a direct one for $\nu_i$ and a seesaw one for $\tilde{\nu}_i$. (Note that $\tilde{\ell}_i$ are not orthogonal.) If large leptonic mixing angles arise from the neutrino sector, $\nu_i$ and $\tilde{\nu}_i$ are expected to be very different, and generically they are not orthogonal. Consequently, the Lagrangian for the light neutrino masses can be written as a sum of three such terms, one from each $\nu_{R_i}$
 \begin{align}
 \label{eq:Lnu}
     {\cal L}_\nu =\frac{1}{2} \sum_i \left( m_i^{\rm dir} \; \nu_i \nu_i - m_i^{\rm ss} \; \tilde{\nu}_i \tilde{\nu}_i \right) + {\rm h.c.} =\frac{1}{2} \sum_i \left(\frac{v_L^2}{v_R^2} \; M_i \;  \nu_i \nu_i
     - \; \frac{y_i^2 v_L^2}{M_i} \; \tilde{\nu}_i \tilde{\nu}_i \right) + {\rm h.c.}.
\end{align}
In parity-symmetric theories, Eq.~\eqref{eq:numassmatrix} demonstrates that  the direct contribution to the neutrino mass, $m_i^{\rm dir} = (v_L^2/v_R^2) M_i$, always contributes to the active neutrino mass, and may dominate over the seesaw contribution, $m_{ij}^{\rm ss}$. A useful numerical parameterization of the direct contribution is
\begin{align}
    \label{eq:dim5Mass}
    m_i^{\rm dir} \simeq 10^{-1} \, {\rm eV} \left(\frac{M_i}{10^9 \, \rm GeV}\right)\left(\frac{6 \times 10^{11} \, \rm GeV}{v_R} \right)^2.
    \end{align}

The lower bounds on $v_R$ derived in this paper follow directly from the equal magnitudes of the $\ell_i \ell_j H_L^2$ and $\bar{\ell}_i \bar{\ell}_j H_R^2$ couplings of (\ref{eq:yukawaNu}), which gives $m_i^{\rm dir} = (v_L^2/v_R^2) M_i$ and (\ref{eq:dim5Mass}).  In conventional LR models, with gauge symmetry breaking from bi-doublets and triplets, the direct contribution to neutrino masses is replaced by a type-II seesaw contribution.  While this contribution is proportional to $(v_L^2/v_R^2) M_i$, the proportionality constant is a free parameter, typically less than unity, so that the bounds on $v_R$ are lost.

\section{Natural Bound on $v_R$ from Thermal Leptogenesis}
\label{sec:vRthermal}
In the early universe, decays of right-handed neutrinos can generate a lepton asymmetry, which is then processed by electroweak sphalerons to give a  cosmological baryon asymmetry. In the case that the lightest right-handed neutrino, $\nu_{R_1}$,
was in the thermal bath before or during
the era of decay, this mechanism is known as thermal leptogenesis. The physics of leptogenesis is the same whether the decaying state is $N_1$ of the augmented Standard Model, (SM + $N$), or $\nu_{R_1}$ of a theory with parity restoration. For simplicity, in both cases we refer to the decaying state as $\nu_{R_1}$. 

After inflation we assume that the universe reheats to a temperature $T_R \gsim M_1$ (except in Sec.~\ref{subsec:nonthermal}) so that thermal production of $\nu_{R_1}$ is not suppressed. We assume that the maximum temperature reached after inflation $T_{\rm max} < v_R$, to avoid domain wall formation at the $SU(2)_R$ phase transition (we remove this constraint in Sec.~\ref{sec:DW} where we add a small $Z_2$-breaking term that makes the wall network unstable). While the EFT of (\ref{eq:yukawaNu}) ensures $M_1 < v_R$, in many theories of inflation $T_{\rm max} \gg T_{RH}$~\cite{Harigaya:2013vwa,Mukaida:2015ria}, so that $v_R \gg M_1$ allows for a wide range of reheating scenarios. Virtual processes involving gauge bosons of mass of order $v_R$ put $\nu_{R_1}$ in thermal equilibrium if \cite{Dror:2020jzy}
\begin{align}
    \label{eq:TRforthermal}
    T_{RH} > 5 \times 10^{10} \, \GEV \left( \frac{v_R}{10^{12} \, \GEV} \right)^{4/3}.
\end{align}
When leptogenesis is close to the strong washout regime, or in strong washout as in Sec.~\ref{subsec:SingletModel}, interactions with
the SM Higgs are sufficient to put 
$\nu_{R_1}$ in thermal equilibrium, so only $T_R > M_1$ is required.

\subsection{Bound on $M_1$}
\label{subsec:M1bound}

Decays of $\nu_{R_1}$ give a yield for the baryon asymmetry of 
\begin{align}
    \label{eq:thermalYB}
    Y_B = \frac{n_B}{s} = \frac{28}{79} \, \epsilon \, \eta\, Y_{\rm therm}  \simeq 10^{-3} \epsilon \, \eta\,  \qquad \text{(Thermal Leptogenesis)},
\end{align}
where $Y_{\rm therm}$ is the thermal yield of $\nu_{R_1}$. The efficiency factor $\eta$ is small when $\nu_{R_1}$ is in thermal equilibrium at decay (the `strong wash-out' regime).
 At the energy scale $M_1$, the Yukawa interaction of $\nu_{R_1}$ is $y_1 \, \tilde{\ell}_1 \nu_{R_1} H_L$, where we use the tilde basis of (\ref{eq:Lyi}).  Integrating out $\nu_{R_1}$ leads to a seesaw mass for $\tilde{\nu}_1$ of size $m_1^{\rm ss} = y_1^2 v_L^2/M_1$. The `strong wash-out' regime is avoided if $m_1^{\rm ss} \lsim 10^{-3}$ eV, in which case $0.3 < \eta < 1$ \cite{Giudice:2003jh}. $\epsilon$ is the lepton asymmetry produced per $\nu_{R_1}$ decay, and results because at 1-loop level $\Gamma(\nu_{R_1} \rightarrow H_L \tilde{\ell}_1) \neq \Gamma(\nu_{R_1} \rightarrow H_L^\dagger \tilde{\ell}_1^\dagger)$.

We begin by assuming that the dominant contribution to the 1-loop diagram for $\epsilon$ involves the exchange of $\nu_{R_2}$,  while the contribution from the exchange of $\nu_{R_3}$ is negligible, giving~\cite{Fukugita:1986hr,Davidson:2008bu}
\begin{align}
    \label{eq:eps0}
    \epsilon = &
    {\rm{Br}}(\nu_{R_1} \rightarrow H_L \ell_1) - {\rm{Br}}(\nu_{R_1} \rightarrow H_L^\dagger \ell_1^\dagger) \nonumber \\
    =& \frac{1}{8\pi}\sum_{j = 2} \frac{\text{ Im}(\mathbf{y}^\dagger {\mathbf{y}})^2_{j1}}{(\mathbf{y}^\dagger {\mathbf{y}})_{11}} g\left(\frac{M_j^2}{M_1^2}\right)=
    \frac{1}{8\pi} \frac{(y_{11}+y_{22})^2}{ y_{11}^2 + |y_{12}|^2} {\rm Im}(y_{12}^2)g(x_2),
\end{align}
where $g(x) = \sqrt{x}((1-x)^{-1}+1-(1+x) 
\ln(1+1/x))$ 
and $x_2 = M_2^2/M_1^2$. $\epsilon$ may be written in terms of the inner product of  $\tilde{\ell}_1$ and $\tilde{\ell}_2$,
\begin{align}
   (\tilde{\ell_1},\tilde{\ell_2})= \frac{(y_{11}+y_{22})y_{12}}{\sqrt{(y_{11}^2 + |y_{12}|^2)(y_{22}^2 + |y_{12}|^2)}} \equiv {\rm cos}\theta_2 e^{i \phi_2},
   \label{eq:l1l2InnerProduct}
\end{align}
so that substituting \eqref{eq:l1l2InnerProduct} into \eqref{eq:eps0} and using the definition of $y_2^2$ gives 
\begin{align}
    \label{eq:eps1}
    \epsilon = \frac{y_2^2}{8\pi} \cos^2 \theta_2 \sin 2\phi_2 \; g(x_2) \, .
\end{align}
Here $\theta_2$ is the angle between the two vectors for $\tilde{\ell}_1$ and $\tilde{\ell}_2$ in flavor space. If they are orthogonal there is no flavor mixing between these two generations in the neutrino sector, and CP violation vanishes.

In the case of degeneracy, $M_2 - M_1 \ll M_2 + M_1$, $g(x) \gg 1$, and we defer this case to section \ref{subsec:degennuR}.  For $M_2$ of order or much greater than $M_1$, a good approximation is $|g(x_2)| \simeq (3/2)M_1/M_2$, giving
\begin{align}
    \label{eq:eps2}
    \epsilon \simeq  \frac{3}{16\pi} \cos^2 \theta_2 \sin 2\phi_2 \; \frac{m_2^{\rm ss} M_1}{v^2}
\end{align}
where $m_2^{\rm ss} = y_2^2 v^2/M_2$ is the seesaw mass generated by the exchange of $\nu_{R_2}$. Requiring that $\epsilon$ is large enough to give the observed baryon asymmetry, $Y_B \simeq 10^{-10}$, and that the seesaw mass from $\nu_{R_2}$ exchange is bounded,
$m_2^{\rm ss} < m_2^{\rm ss \, *}$, leads to a lower bound on $M_1$
\begin{align}
    \label{eq:M1bound}
    M_1 \gsim  \; \frac{6 \times  10^8 \; \GEV}{\eta A_2} \;  \left( \frac{0.05 \EV}{m_2^{\rm ss \, *}}\right), \, \hspace{0.25in} {\rm with} \hspace{0.25in}  A_2 = \cos^2 \theta_2 \sin 2\phi_2.
\end{align}
The bound becomes stronger as $\eta$, $\cos^2 \theta_2$ or $\sin 2\phi_2$ are taken less than unity.  

The contribution to $\epsilon$ from $\nu_{R_3}$ exchange takes the same form as that from $\nu_{R_2}$ exchange, so that together they yield
\begin{align}
    \label{eq:eps23}
    \epsilon \simeq  \frac{3}{16\pi} (A_2 m_2^{\rm ss} + A_3 m_3^{\rm ss}) \; \frac{M_1}{v^2}, 
\end{align}
where $m_3^{\rm ss} = y_3^2 v^2/M_3$, $A_3 = \cos^2 \theta_3 \sin 2\phi_3$ and $\theta_3$ is the angle between $\tilde{\ell}_1$ and $\tilde{\ell}_3$. We have also taken $M_3$ to be of order or much greater than $M_1$.

Taking $m_{2,3}^{\rm ss} \, \lesssim \, m_{2,3}^{\rm ss \, *}$, the lower bound on $M_1$ becomes
\begin{align}
    \label{eq:M1bound}
    M_1 \; \gsim  \; \frac{6 \times  10^8 \; \GEV}{\eta (A_2 + A_3)} \;  \left( \frac{0.05 \, \EV}{m_{2,3}^{\rm ss \, *}}\right) > \; 3 \times  10^8 \; \GEV  \left( \frac{0.05 \, \EV}{m_{2,3}^{\rm ss \, *}}\right) . \,
\end{align}
Can this bound on the lightest right-handed neutrino  be weakened by increasing ${m_{2,3}^{\rm ss \, *}}$ above 0.05 eV? The cosmological limit on the sum of the neutrino masses \cite{DES:2021wwk} is now so severe that the three neutrinos cannot be made almost degenerate with masses larger than 0.05 eV.  {\it Significant weakening requires a fine-tuning to force a cancellation among different contributions to a light neutrino mass eigenvalue.} In this paper we do not allow such tunings and hence study the consequences of the approximate naturalness bound
\begin{align}
    \label{eq:M1boundnat}
    M_1 \; \gsim  \;  10^9 \; \GEV. \,
\end{align}
This bound on $M_1$ applies in (SM$+ N$) and in all parity symmetric theories where neutrino masses result from the operators of (\ref{eq:yukawaNu}). The precise numerical value of the bound depends on the amount of cancellation or tuning, parameterized by ($m_{2.3}^{\rm ss \, *}/0.05$ eV) in Eq. \eqref{eq:M1bound}. 

In the `strong wash-out' regime the efficiency $\eta$ is much less than unity, strengthening this bound on $M_1$. The bound on $M_1$ in Figs.~\ref{fig:naturalnessPlot}, \ref{fig:naturalnessPlotdegen} and \ref{fig:naturalnessPlotSinglet} are for weak washout with $\eta \simeq 1$ in order to show the most optimistic lower limits on $M_1$ and $v_R$ and to demonstrate such limits are still severe. After the lepton asymmetry is created by $\nu_{R_1}$ decay, removal of the asymmetry via $\nu_{R_1}$ production is small provided $m_1^{\rm ss} \lsim 10^{-3} \, {\rm eV}$; for larger values of $m_1^{\rm ss}$, the bound on $M_1$ is strengthened by approximately $(m_1^{\rm ss} / 10^{-3} \, \mbox{eV})^{1.16}$ \cite{Giudice:2003jh}. Recall that (\ref{eq:M1bound}) applies even if $M_{2,3} \gg M_1$, in which case washout by $\nu_{R_{2,3}}$ production is negligible. Even in the case that $M_{2}$ is comparable to $M_1$, and $m_2^{\rm ss} \simeq 0.05 \, {\rm eV}$ to maximize production of the lepton asymmetry, strong washout via $\nu_{R_2}$ production can be avoided by reducing $m_1^{\rm ss}$ well below $10^{-3} \, {\rm eV}$ so that $\nu_{R_1}$ decays at a temperature well below $M_1 \sim M_2$ (but before the universe becomes matter dominated by $\nu_{R_1}$).  In this case, $T_{RH}$ should satisfy (\ref{eq:TRforthermal}), as the Yukawa coupling of $\nu_{R_1}$ is too small to put it in thermal equilibrium.
$\nu_{R_1}$ can also decay via $W_R$ exchange, but we find that this decay mode is negligible and does not reduce the efficiency of leptogenesis.

We stress that this bound results from thermal leptogenesis without degeneracy among the $\nu_{R_i}$; cases with a non-thermal initial abundance or with degeneracy are discussed in sections \ref{subsec:nonthermal} and \ref{subsec:degennuR}. This naturalness bound applies even if there are additional contributions to the light neutrino masses coming directly from dimension-5 operators.  This necessarily occurs in the parity symmetric theories, but is also expected in (SM + $N$) which, after all, is just an effective field theory. Finally, we note that in (SM + $N$) this bound is not the rigorous Davidson-Ibarra bound \cite{Davidson:2002qv}, derived for $M_{2,3} \gg M_1$; rather, for moderate $M_{2,3}$ it can be violated if there are unnatural cancellations in the light neutrino mass matrix \cite{Hambye:2003rt}.

A value of $M_1$ of order $10^{9}$ GeV only results for leptogenesis with weak washout and if the relevant angles in $A_2$ or $A_3$ are order unity, otherwise the naturalness bound on $M_1$ can be orders of magnitude more severe.
 
As an example of the tuning required to avoid \eqref{eq:M1boundnat}, consider the case with only seesaw neutrino masses, which is possible in both (SM + $N$) and theories with parity restoration with sufficiently high $v_R$. Leptogenesis with weak washout, $m_1^{\rm ss} \lsim 10^{-3} \, {\rm eV}$, implies that only the seesaw masses from $\nu_{R_{2,3}}$ are relevant for current neutrino mass observations. One neutrino can be taken as essentially massless and the mass matrix for the two heavy states can be put in the form
\begin{align}
\label{eq:mnu2by2}
 {\bf m_\nu} =
 \begin{pmatrix}
     \sin^2 \theta \, m_2^{\rm ss} & \cos \theta \sin \theta \, m_2^{\rm ss} \\ \cos \theta \sin \theta\,  m_2^{\rm ss} \;\;\;\; & e^{2i\phi} \, m_3^{\rm ss}  + \cos^2 \theta \,m_2^{\rm ss} 
    \end{pmatrix} \,
\end{align}
where the angles $\theta$ and $\phi$ specify the orientation between the two vectors for $\tilde{\ell}_2$ and $\tilde{\ell}_3$ in flavor space and are not determined by leptogenesis. If $\theta \neq 0$ the mass eigenvalues are not $m_{2,3}^{\rm ss}$ because $\nu_{R_{2,3}}$ couple to different combinations of light neutrinos.  If $\phi=0$, $m_{2,3}^{\rm ss} \gg 0.05$ eV is excluded: a cancellation between them is not possible as they are both positive.  On the other hand for a range of $\phi$ a cancellation is possible. For example,  $m_{2,3}^{\rm ss} \gg 0.05$ eV is possible if $\phi = \pi/2$, giving $m_{\rm atm} \simeq m_3^{\rm ss} - m_2^{\rm ss}$ and $m_\odot \simeq \sin^2 \theta \; (m_{2,3}^{\rm ss})^2 / m_{\rm atm}$. Here $m_{\rm atm} \simeq 0.05$ eV and $m_{\odot} \simeq 0.01$ eV describe the mass splittings for atmospheric and solar oscillations. In this paper we explore the consequences of avoiding such tunings.

\subsection{Bound on $v_R$}
\label{subsec:vRbound}

In parity symmetric models the right-handed neutrinos are massless until $SU(2)_R$ is broken, implying that $v_R \gsim M_+$, the mass of the heaviest right-handed neutrino, which could be $M_2$ or $M_3$. Hence, thermal leptogenesis, via \eqref{eq:M1boundnat}, already implies that $v_R \gsim 10^9$ GeV. However, there is a much stronger bound. Parity invariance of the dimension-5 operators implies a direct contribution to the light neutrino masses, $m_i^{\rm dir}$, which is $v_L^2/v_R^2$ times the masses of $\nu_{R_i}$, as shown in Eq. \eqref{eq:dim5Mass}. In the absence of fine-tuning in the light neutrino mass matrix, $ m_+^{\rm dir} \lsim m_{+}^{{\rm dir}*}$ so that 
\begin{align}
    \label{eq:vRbound1}
 v_R \;\; \gsim \;\;  10^{12} \, {\rm GeV}   \left(\frac{M_+}{10^9 \, {\rm GeV}}\right)^{\frac{1}{2}} \left( \frac{0.05 \, {\rm eV}}{m_{+}^{{\rm dir}*}} \right)^{\frac{1}{2}}.
 \end{align}
Hence, the thermal leptogenesis bound of (\ref{eq:M1bound}) leads to a numerical bound on $v_R$ of
\begin{align}
    \label{eq:vRbound2}
 v_R \; \; \gsim \; \;  10^{12} \, {\rm GeV} \left(\frac{M_+}{M_1} \right)^{\frac{1}{2}} \left(\frac{1}{(A_2 + A_3) \eta }\right)^{\frac{1}{2}} \left( \frac{0.05 \, {\rm eV}}{m_{+}^{{\rm dir}*}} \right)^{\frac{1}{2}}  \left( \frac{0.05 \, {\rm eV}}{m_{2,3}^{\rm ss \, *}} \right)^{\frac{1}{2}} \, .
    \end{align}
This bound is applicable only when there is no high degeneracy between $M_1$ and $M_2$ or $M_3$. This case is studied in Sec.~\ref{subsec:degennuR} and is found to weaken the bound by up to three orders of magnitude. 

The lowest values for $v_R$ are obtained in weak washout, with $\eta \simeq 1$. Washout is negligible, for example, when $m_1^{\rm ss} \sim 10^{-3}$ eV, and $M_{2,3}/M_1 \gtrsim 4$ so that  $\nu_{R_{2,3}}$ production is Boltzmann suppressed at the time of $\nu_{R_1}$ decay, making $\nu_{R_{2,3}}$ back-reactions small. The lowest value of $v_R$ occurs when: $M_{2,3}$ is comparable to $M_1$; $m_{2,3}^{\rm ss} \simeq 0.05 \, {\rm eV}$ to maximize production of the lepton asymmetry; and strong washout via $\nu_{R_{2,3}}$ production is avoided by reducing $m_1^{\rm ss}$ well below $10^{-3} \, {\rm eV}$, so that $\nu_{R_1}$ decays at a temperature well below $M_1 \sim M_2 \sim M_3$. 
If washout is important, the bound on $v_R$ becomes stronger than \eqref{eq:vRbound2}. 

Remarkably, from Fig.~\ref{fig:vpPrediction} we see that the central value of the prediction for $v_R$ from the Higgs Parity mechanism is of order $10^{12}$ GeV, although the 3$\sigma$ range spans several orders of magnitude.  For lower values of the top quark mass the above bound on $v_R$ is easily satisfied. However, larger values are inconsistent with this bound, and would provide a strong motivation to modify the minimal leptogenesis scheme of the previous sub-section. 
We stress that the bound on $v_R$ is strengthened as the right-handed neutrinos become more hierarchical in mass; if future measurements of $m_t$ are high, consistency will limit this hierarchy. 
Thermal leptogenesis with small angles in $A_2$ or $A_3$, strengthens the  bound on $v_R$, but more mildly than for the bound on $M_1$.  The bounds of (\ref{eq:M1bound}) and (\ref{eq:vRbound2}) are both weakened linearly if the relevant $m^*$ are taken larger than 0.05 eV.  For example, if cancellations of a factor of three occur between differing contributions to a light neutrino mass eigenvalue, then $m^* = 0.15$ eV and the lower limits on $M_1$ and $v_R$ are both lowered by a factor of 3.

\begin{figure}[tb]
    \centering
    \includegraphics[width=.8\textwidth]{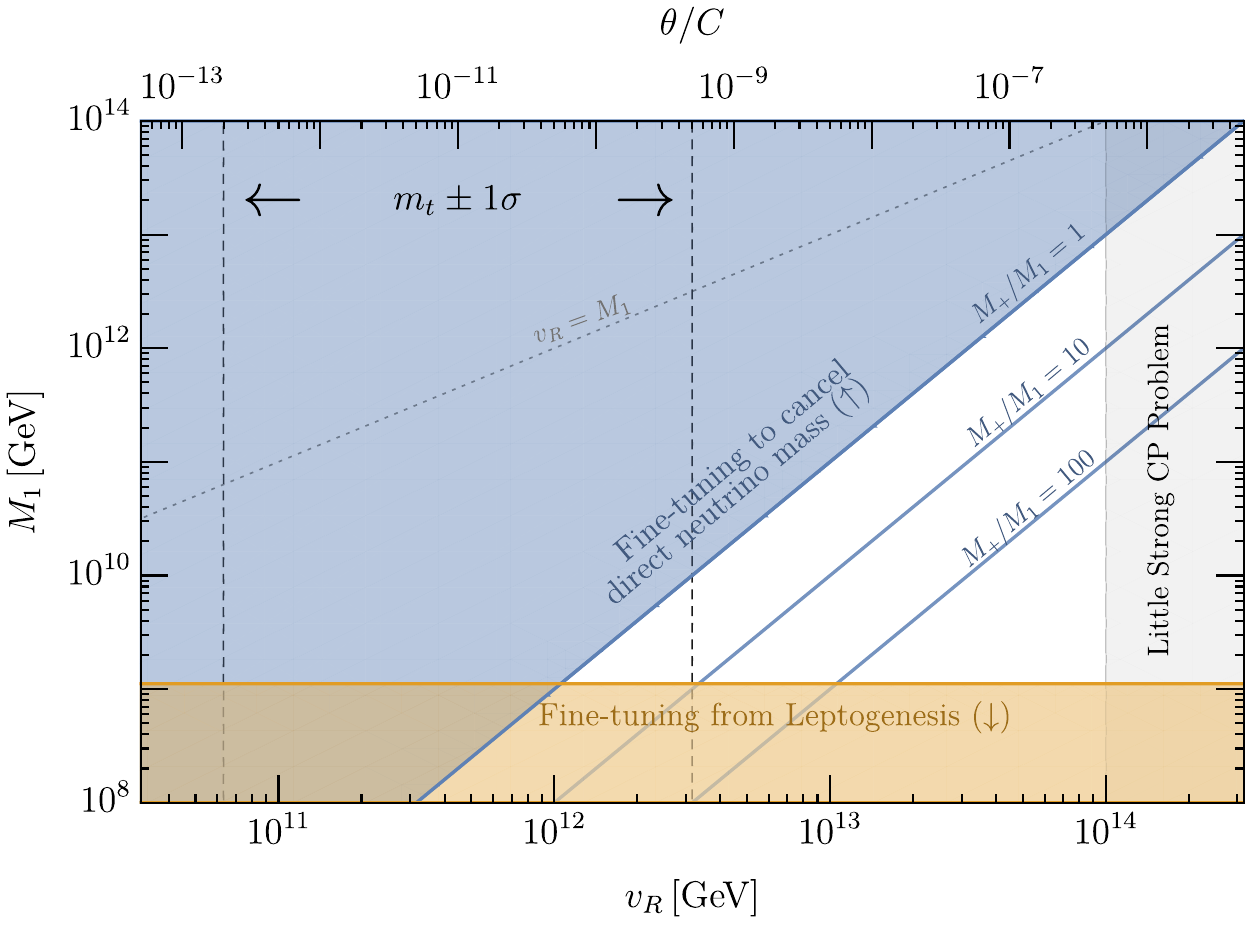} 
    \caption{
    Lower bounds on $M_1$ and $v_R$ in minimal theories where parity solves the strong CP problem and $\nu_{R_1}$ decay gives successful leptogenesis. Neutrino masses arise from the dimension-5 operators of (\ref{eq:yukawaNu}) without fine-tuning. The horizontal bound is from requiring the observed baryon asymmetry to arise from thermal leptogenesis without degeneracy of $\nu_R$. The sloped bound arises from limits on direct contributions to light neutrino masses, for three illustrative values of $M_+/M_1$. In the shaded blue and orange regions, fine-tuning between contributions to the neutrino masses increases with distance from the natural unshaded region. At values of $v_R$ larger than $10^{14}$ GeV the strong CP problem re-emerges from (\ref{eq:dim6}), as illustrated by the gray shading and the top axis. The region above the dashed line for $M_1=v_R$ is unphysical in the EFT of (\ref{eq:yukawaNu}).}
    \label{fig:naturalnessPlot}
\end{figure}

The bounds of (\ref{eq:M1bound}) on $M_1$ and (\ref{eq:vRbound1}) on $v_R$ are shown in Fig.~\ref{fig:naturalnessPlot} in the $(v_R, M_1)$ plane. As $v_R$ is increased above about $10^{14}$ GeV, the strong CP problem re-emerges. Hence, in the unshaded region, which successfully accounts for both neutrino masses and the cosmological baryon asymmetry, there is a good possibility that the operator of (\ref{eq:dim6}) will give a positive signal for the neutron electron dipole moment in current experiments, providing the coefficient $C$ is not small, $C > 0.01$. The unshaded region overlaps the region of $v_R$ predicted by Higgs Parity, illustrated by the vertical dashed lines showing the range of $v_R$ allowed by the current $1 \sigma$ uncertainty in $m_t$. As uncertainties on $m_t$ and $\alpha_s$ are reduced, it will be interesting to discover whether consistency between Higgs Parity and thermal leptogenesis is maintained. 

In the `strong wash-out' regime, the bound (\ref{eq:vRbound2}) on $v_R$ strengthens, as it is proportional to  $1/\sqrt{\eta} \simeq (m_1^{\rm ss} / 0.5 \times 10^{-3} \mbox{eV})^{0.58}$ \cite{Giudice:2003jh}, enlarging the blue region of Fig.~\ref{fig:naturalnessPlot}. For example, for $m_1^{\rm ss} = 0.003 \, \mbox{eV}$ this is about a factor 3.  Entering the strong washout regime removes the bound (\ref{eq:TRforthermal}) on $T_R$ for thermal leptogenesis -- it is only necessary that $T_R \gsim M_1$.  This opens up more parameter space for the theory of reheating after inflation, which should now satisfy $v_R > T_{\rm{max}} > T_R > M_1$.

The lower bound (\ref{eq:vRbound2}) on $v_R$ follows directly from the equal magnitudes of the $\ell_i \ell_j H_L^2$ and $\bar{\ell}_i \bar{\ell}_j H_R^2$ couplings of (\ref{eq:yukawaNu}).  In conventional LR models, electroweak symmetry is broken by bi-doublets, $\Phi$, and triplets, $\Delta_{L,R}$. 
The $\bar{\nu}\bar{\nu}$ mass term arises from the vev of $\Delta_R$, and there is a type-II seesaw $\nu \nu$ contribution from $\Delta_L$ exchange.  If the mass of $\Delta_L$, $M_{\Delta_L}$, is of order $v_R$, the ratio of these two mass terms is proportional to $v_L^2/v_R^2$. However, it is also proportional to $\lambda_{LR}$, the quartic coupling for the $\Phi^\dagger \Phi \Delta_L \Delta_R$ operator \cite{Dror:2020jzy} 
\begin{align}
    \label{eq:convLR}
 r \; = \; \left( \frac{\rm{type} \; II \; \rm{seesaw} \;\; \nu \nu \;\; \rm{mass} }{\bar{\nu}\bar{\nu} \;\; \rm{mass} } \right) \; = \; \lambda_{LR} \; \frac{ v_L^2}{M_{\Delta_L}^2}.
    \end{align}
If $\lambda_{LR} \sim 1$, the order of magnitude of the bound survives; but if $\lambda_{LR} \ll 1$, the bound is considerably weakened by a factor of $\sqrt{r}$.  The bound would be strengthened for $m \ll v_R$, but this requires fine-tuning.

\subsection{The $\nu_R$ Spectrum}
\label{subsec:nuRspectrum}

In this paper we define $\nu_{R_1}$ to be the right-handed neutrino whose decays yield leptogenesis.  In the previous sections, for convenience, we also took $\nu_{R_1}$ to be the lightest right-handed neutrino, but this is unnecessary. For example, we can take leptogenesis to arise from virtual $\nu_{R_2}$ exchange in $\nu_{R_1}$ decays, with $M_2 \geq M_1$, and take $M_3$ much larger or much smaller than $M_1$ as long as it does not upset the leptogenesis mechanism. This requires that $\nu_{R_3}$ is much lighter and decays while relativistic, or decays soon after it becomes non-relativistic; in either scenario it will not washout the lepton asymmetry previously created by $\nu_{R_1}$ decay.  However, if $\nu_{R_3}$ comes to dominate the energy density of the universe, then its decays will dilute the lepton asymmetry, and we do not allow this case.  It is even possible that $\nu_{R_3}$ is in the ($\sim$2-100) keV mass range and is sufficiently stable to be dark matter \cite{Dunsky:2020dhn}. Finally, even if $W_R$ exchange led to an initial thermal abundance of right-handed neutrinos, it it still possible that $\nu_{R_1}$ is sufficiently heavy and long-lived to come to dominate the energy density of the universe before decaying, thereby diluting $\nu_{R_3}$ to the observed dark matter abundance.

\section{Relaxing the Leptogenesis Bound on $v_R$}
\label{sec:vRrelax}
The lower bound on the scale of parity violation derived in the last section from thermal leptogenesis, $v_R >10^{12}$ GeV, can be weakened by a factor $1/ \sqrt{E}$, when the baryon asymmetry from leptogenesis is enhanced by a factor $E$. 
There are two well-known ways of enhancing leptogenesis. In the first a cosmological mechanism is introduced to increase the yield of $\nu_{R_1}$ above thermal. In the second, the lepton asymmetry per $\nu_{R_1}$ decay is increased by having degeneracy among the $\nu_{R_i}$, which can result from an approximate symmetry. In the first two subsections we study how each of these affects the lower bound on $v_R$ in theories with parity restoration.  In the final subsection, we investigate how the bounds on $v_R$ are modified when the dimension-five operators for neutrino masses arise from a very simple theory, the Radiative Singlet Model, where seesaw and direct contributions to light neutrino masses cancel at tree-level.


\subsection{Non-Thermal Leptogenesis}
\label{subsec:nonthermal}

In this subsection, we study the possibility that before decaying, the $\nu_{R_1}$ are not in thermal equilibrium and have a number density greater than the thermal value.
We show that this case can relax the bound on $v_R$ only by about an order of magnitude.

Consider a cosmological era with the energy density of the universe dominated by $\nu_{R_1}$. At the end of this era, when $\nu_{R_1}$ decay, we take the $\nu_{R_1}$ number density to be $n_D$ and their momentum distribution to be peaked around $p_D$, corresponding to a typical energy $E_D = \sqrt{M_1^2 + p_D^2}$.
If the decay rate of $\nu_{R_1}$ at rest is $\Gamma_1$, the Hubble parameter at decay is
\begin{align}
    \label{eq:HD}
    H_D \simeq \Gamma_1 \frac{M_1}{E_D}. 
\end{align}
The decay products, $\ell,H_L$ and their anti-particles, reheat the universe via gauge interactions to a temperature $T_{RH}$, given by
\begin{align}
    \label{eq:TRH}
    \frac{\pi^2}{30} g_* T_{RH}^4  \simeq \rho_D \simeq n_D E_D. 
\end{align}
At the same time, the $\nu_{R_1}$ decays produce a lepton number density $\epsilon n_D$ which, in the resulting thermal bath, becomes a baryon yield
\begin{align}
    \label{eq:nonthermYB}
    Y_{B,\rm non-therm} = \frac{28}{79} \,  \epsilon \eta  \left(\frac{3}{4}\frac{T_{\rm RH}}{E_D}\right) \,. \quad \text{(Non-Thermal Leptogenesis)}
\end{align}
Comparing with the result (\ref{eq:thermalYB}) for thermal leptogenesis, there is an enhancement factor
\begin{align}
    \label{eq:YBcomparison}
 F \equiv   \frac{Y_{B,\rm non-therm}(\eta)}{Y_{B,\rm therm}(\eta = 1)} \simeq  200 \frac{T_{RH}}{E_D} \eta,
\end{align}
for equal values of $\epsilon$.  This ratio can exceed unity, and hence non-thermal leptogenesis can produce the observed baryon asymmetry with a smaller $\epsilon$ and therefore a smaller $M_1$ than in the thermal case.  However, the size of this ratio and its dependence on $E_D$ is limited by the Pauli exclusion principle and by washout. 

The Pauli exclusion principle limits the number density of $\nu_{R_1}$ at decay: $n_D \lesssim p_D^3/\pi^2$. Using this in (\ref{eq:TRH}) gives
\begin{align}
\label{eq:PEP}
 \frac{T_{RH}}{E_D} \; \lesssim \; \frac{1}{\pi} \left(\frac{p_D}{E_D}\right)^{3/4}.
\end{align}
Thus the enhancement factor, $F$, is below unity in the non-relativistic region when $p_D < M_1/250$, rises as $p_D$ is increased, but is no more than $\sim 60$ in the relativistic regime. 

After reheat there is the possibility that scattering of the leptons $\ell$ via the Yukawa coupling to $\nu_{R_1}$ will washout the lepton asymmetry. When $p_D \lesssim M_1$, (\ref{eq:PEP}) gives $T_{RH} < M_1$, giving the weak washout condition 
\begin{align}
\label{eq:washoutNT}
 \Gamma_{\rm washout}  \; \simeq \; \Gamma_1 \frac{T_{RH}}{M_1} \; e^{-(M_1/T_{RH})} \; \lesssim \; H_D \; \simeq \; \Gamma_1 \frac{M_1}{E_D}.
\end{align}
which is satisfied for all $p_D \lesssim M_1$. As $p_D$ approaches the relativistic region, $F$ rises to near 60.
In the relativistic region, with $T_{RH} > M_1$,  the weak washout condition is the same as for thermal leptogenesis, $m_1^{\rm ss} \lsim 10^{-3}$ eV.  However, such high values for $T_{RH}$ require a rapid decay rate for $\nu_{R_1}$, and are inconsistent with this weak washout condition.  Therefore, for relativistic $\nu_{R_1}$ decays the enhancement factor is $F \simeq 60 \eta$ with $\eta \simeq {\cal O} (10) (M_1 / E_D)^{3.5}$. 
Hence,  by the same logic as Sec.~\ref{subsec:vRbound}, in non-thermal leptogenesis with $p_D \sim M_1$, the limit on $v_R$ can be weakened by $\sqrt{F}$ compared to thermal leptogenesis to
\begin{align}
\label{eq:vRboundnonth}
 v_R \;\; \gsim \;\;  10^{11} \, {\rm GeV}.
\end{align}
When the bound is saturated, the reheat temperature is around $10^6$ GeV.

Leptogenesis with nearly degenerate right-handed neutrinos can indeed be achieved when the inflaton, $\phi$, decays dominantly by $\phi \rightarrow \nu_{R_1} \nu_{R_1}$. If the inflaton number density is sufficiently high when Hubble is of order the inflaton decay rate, $\Gamma_\phi$, the decays are Pauli-blocked. This can be seen by noting the inflaton energy density $\rho_\phi = m_\phi n_\phi = 3H^2/8\pi G$, so that the the faster the $\phi$ decay rate is (occurring at $\Gamma_\phi \simeq H$), the greater the inflaton number density, $n_\phi$, is. If $\Gamma_\phi  \simeq H \gg m_\phi^2/ 4 \pi M_{Pl}$, the inflaton number density at decay is $n_\phi \gg m_\phi^3$, which would generate a number density of $\nu_{R_1}$ much greater than the Pauli degenerate limit of $n_{\rm PD} \simeq (m_\phi/2)^3/\pi^2$. This is forbidden and so $\phi$ decays must occur over a long period until $n_\phi$ drops below $n_{\rm PD}$. During this period, $\phi$ produces a nearly degenerate Fermi gas of $\nu_{R_1}$ which are continually red-shifted by the expansion of the universe. The phase space of the $\nu_{R_1}$ is filled until the last inflatons to decay populate states near the Fermi surface, which dominate the energy density of the distribution. The non-thermal but relativistic $\nu_{R_1}$ then redshift with the expansion until decaying. Achieving the largest possible enhancement factor over thermal leptogenesis, $F \simeq 60$, requires that $\nu_{R_1}$ be at least semi-relativistic and in the weak washout regime. With an initial momentum of $m_\phi/2$ there can therefore be no more than a factor of $(m_\phi/2)/(M_1)$ in the redshift between $\nu_{R_1}$ production from the inflaton and $\nu_{R_1}$ decay. This maximum possible redshift implies the minimum energy density of $\nu_{R_1}$ at decay is $\rho_D \gtrsim m_\phi n_{\rm PD}(2M_1/m_\phi)^4 \approx M_1^4/\pi^2$. Equivalently, the $\nu_{R_1}$ decay rate must be 
\begin{align}
    \Gamma_1 \approx \frac{M_1^2}{M_{Pl}} \,.
\end{align}
for the lowest limit on $v_R$ to be achieved, Eq. \eqref{eq:vRboundnonth}. 
See~\cite{Lazarides:1990huy,Asaka:1999yd} for leptogenesis from the inflaton decay with a more generic parameter space. 

\subsection{Degenerate Right-Handed Neutrinos}
\label{subsec:degennuR}
In this subsection, we show how increasing the lepton asymmetry per decay, $\epsilon$, via degenerate right-handed neutrinos can relax the natural lower bound on $v_R$ by three orders of magnitude to $10^{9}$ GeV.

Consider the case where at least two right-handed neutrinos are so degenerate that their mass difference $\Delta M$, is much smaller than their mass, $M_1 \simeq M_2 \simeq M$.  The lepton asymmetry generated by the decay of $\nu_{R_1}$ is then enhanced by the large $g(x)$ factor in Eq. \eqref{eq:eps1}, which takes the form of
\begin{align}
    \label{eq:gDegeneracy}
    \left|g\left(\frac{M_2^2}{M_1^2}\right) \right| \simeq \frac{1}{2}\frac{M}{|\Delta M|} \gg 1
\end{align}
for $M_1 \simeq M_2 \simeq M$.
\footnote{Eq. \eqref{eq:gDegeneracy} breaks down when $\nu_{R_1}$ and $\nu_{R_2}$ are so degenerate that $|\Delta M| \lesssim \Gamma_2$, where $\Gamma_2$ is the decay rate of $\nu_{R_2}$. If this occurs, $g(x)$ is not as enhanced as \eqref{eq:gDegeneracy} and thus the lower limit on $v_R$ is greater than $10^9$ GeV.}

Such a degeneracy between $M_1$ and $M_2$ can arise, for example, if there exists a discrete symmetry in the lepton number violating operator of \eqref{eq:yukawaNu} 
\begin{align}
    \label{eq:dissym}
    \ell_1 \leftrightarrow \ell_2
    \qquad \ell_1 \leftrightarrow -\ell_1 \, ,
\end{align}
and similarly on the $\bar{l}$ fields to maintain parity, so that $c_{ij}$ is diagonal with $c_{11} = c_{22}$ giving $M_1 = M_2 = |c_{11}|^2 v_R^2/2M$ \cite{Dunsky:2020dhn}. This symmetry cannot be imposed on the lepton number conserving operator of \eqref{eq:yukawaNu} since if $b_{ij}$ is also diagonal there is no flavor violation for leptogenesis. Similarly, it cannot be imposed on the operators generating charged lepton masses, as there is no degeneracy in the charged lepton spectrum. In the underlying theory of lepton flavor, there must be different patterns of symmetry breaking in the lepton number preserving and lepton number violating sectors.

\begin{figure}[tb]
    \centering
    \includegraphics[width=.8\textwidth]{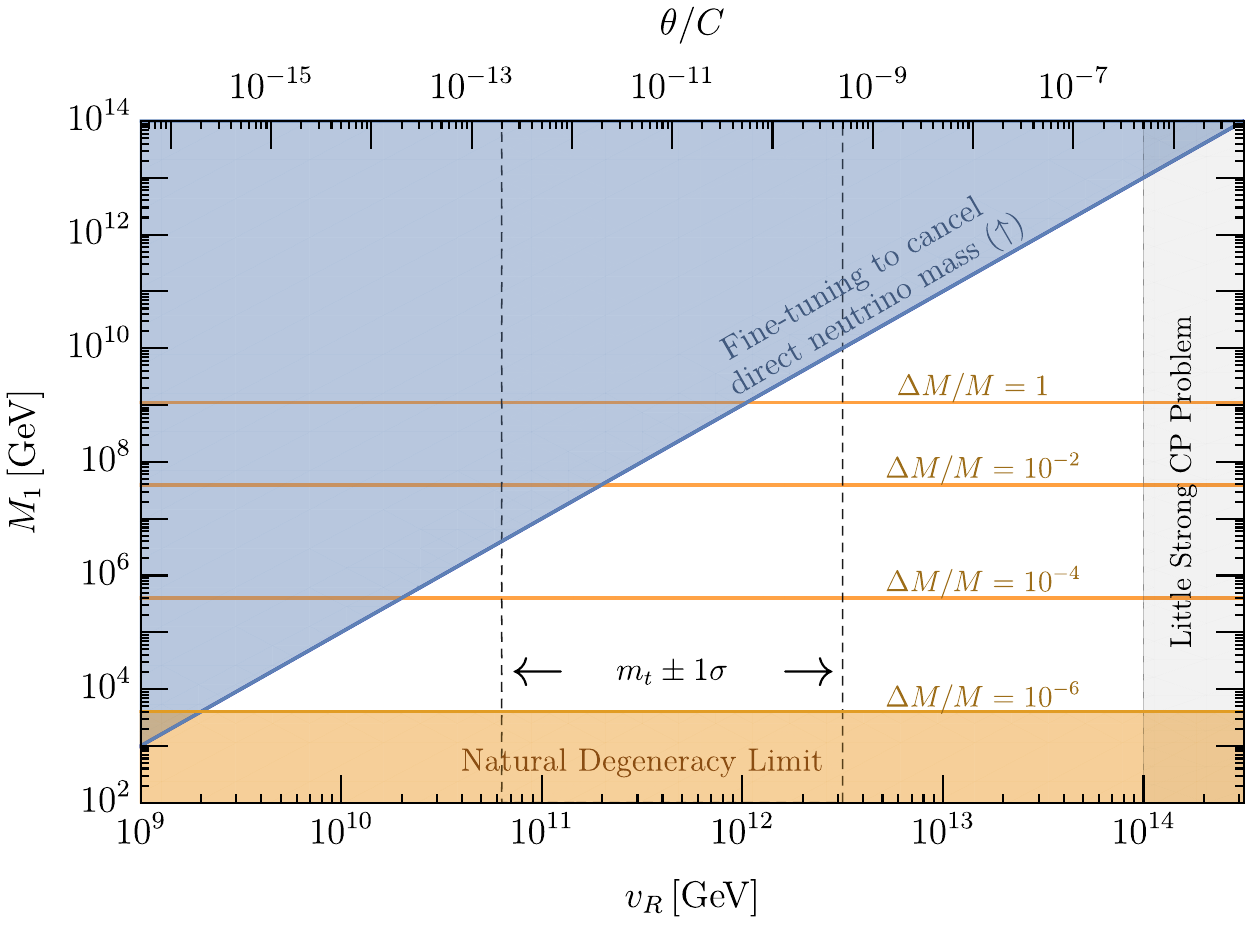} 
    \caption{Lower bounds on $M_1$ and $v_R$ in theories of parity restoration, with minimal scalars and no fine-tuning in neutrino masses. The horizontal bound is from requiring the observed baryon asymmetry to arise from thermal leptogenesis; it is shown for several values of the degree of degeneracy $\Delta M/M$ between the decaying and virtual $\nu_R$. The case without degeneracy, $\Delta M/M \approx 1$, gives the bound of Fig.~\ref{fig:naturalnessPlot}. As $\Delta M/M$
    is decreased to its smallest natural value of $10^{-6}$, the bound on $M_1$ decreases linearly.  As indicated by the gray shaded region, the strong CP problem re-emerges at values of $v_R$ larger than $10^{14}$ GeV, as discussed around \ref{eq:dim6}.}
    \label{fig:naturalnessPlotdegen}
\end{figure}

How small can $\Delta M/M$ be? In Appendix \ref{app:degeneracy}, we demonstrate that the charged fermion Yukawa operators necessarily generate a wavefunction renormalization for $\nu_{R_{1,2}}$ and hence a mass splitting $\Delta M$ that is at least of order $y_\tau^2/8\pi \sim 10^{-6}$. Eq. \eqref{eq:gDegeneracy}  implies  then that $g(x)_{\rm max} \simeq 5 \times 10^5$ is the largest natural value in any LR theory. The naturalness bound of \eqref{eq:M1bound} is greatly weakened and becomes
\begin{align}
    \label{eq:M1bounddeg}
    M_1 \; \gsim  \; \frac{2 \times 10^3 \; \GEV}{A_2} \left(\frac{5 \times 10^5}{g(x)_{\rm max}} \right)
    \left( \frac{0.05 \, \EV}{m_{2}^{\rm ss \, *}}\right)
    \hspace{.25in}  \,
    \end{align}
for weak washout.\footnote{This result follows from the lepton asymmetry produced by $\nu_{R_1}$ decay; a comparable lepton asymmetry from $\nu_{R_2}$ decay could weaken the bound by up to a factor of 2.} The horizontal orange lines in Fig.~\ref{fig:naturalnessPlotdegen} show the limit on $M_1$ for several values of the degeneracy $\Delta M/M$. Note that, since $\nu_{R_1}$ and $\nu_{R_2}$ are degenerate, for $T_{RH} \sim M_1$ a value of $m_{1,2}^{\rm ss}$ as large as 0.05 eV will lead to strong washout, strengthening the bound on $M_1$ by a factor of 50. However, strong washout can be avoided by taking $T_{RH}$ somewhat less than $M_1$ so that washout through $\nu_{R_2}$ production is exponentially suppressed. 

The blue curve of Fig \ref{fig:naturalnessPlotdegen} shows the lower bound on $v_R$ from the direct contribution to the light neutrino mass as a function of $M_1$; it is (\ref{eq:vRbound1}) with $M_i$ replaced by $M_1$. 
The reduction in the lower bound on $M_1$ from leptogenesis by a factor of order $10^6$, decreases the naturalness bound of \eqref{eq:vRbound2} on $v_R$ by $10^3$ to
\begin{align}
\label{eq:vRbounddeg}
    v_R \;\; \gsim \;\; \frac{10^{9} \, {\rm GeV}}{(A_2)^{1/2}}  \left(\frac{5 \times 10^5}{g(x)_{\rm max}} \;\;
    \frac{M_+}{M_1} \;\;
    \frac{0.05 \, {\rm eV}}{m_{+}^{{\rm dir}*}} \right)^{\frac{1}{2}} 
     \left(\frac{0.05 \, \EV} {m_{2}^{\rm ss}}\right)^{\frac{1}{2}}
\end{align}
for weak washout. Taking $M_+ = M_3 \sim M_1$, this bound corresponds to the intersection of the blue line with the orange lines in Fig \ref{fig:naturalnessPlotdegen}.

It is interesting to note that for such highly degenerate $\nu_{R_{1,2}}$ the direct contributions to the light neutrino masses become equal, $m_{1,2}^{\rm dir} = (v_L^2/v_R^2) M_{1,2}$.
\footnote{When $\nu_{R_2}$ and $\nu_{R_3}$ are nearly degenerate, it is possible to weaken the bound on $M_1$ by a factor of 2 and the bound on $v_R$ by a factor of $\sqrt{2}$ if the following two conditions are met: 1) $y_1^2 \simeq y_2^2$, 2) Their CP phases are opposite, $\phi_1 = -\phi_2$, so as to cancel the negative sign picked up by $g(M_2^2/M_1^2) \simeq - g(M_1^2/M_2^2)$.}

In the Mirror theory, a large degeneracy allows for the possibility that the reheating temperature required for leptogenesis can be far below the mass of the lightest mirror quark
 \begin{align}
 \label{eq:U1^2mirror}
     T_{\rm RH} \simeq M_1 \ll m_{u'} \simeq 10^{-5} v_R.
\end{align}
In this case, there may be negligible production of $u'$ after inflation, allowing $u'$ to be stable and still below terrestrial bounds on fractionally charged particles \cite{Dunsky:2018mqs}. This allows an additional $U(1)$ in the theory so that the Mirror theory can have electroweak gauge group $SU(2)_L \times U(1)_Y \times SU(2)_R \times U(1)'_{Y'}$ as in \cite{Dunsky:2019api, Bonnefoy:2023afx}.

\subsection{Radiative Singlet Model}
\label{subsec:SingletModel}
In our parity symmetric theory, neutrino masses are described by the dimension-5 operators of (\ref{eq:yukawaNu}), in both LR and Mirror versions.  The physics of neutrino masses and leptogenesis depends on two independent flavor matrices, $b$ and $c$. Significant cancellations between different contributions to the light neutrino mass matrix can only arise from fine-tuning the parameters of these matrices, leading to the naturalness bounds discussed so far in this paper. However, cancellations between seesaw and direct contributions to neutrino masses occur naturally in one of the simplest UV completions of this EFT~\cite{Hall:2019qwx}, which we call the Radiative Singlet Model. In this theory the $b$ and $c$ matrices are correlated, resulting in all three active neutrino masses vanishing at tree-level. In this section, we study whether such cancellations lead to a relaxation of the naturalness bound on $v_R$, finding that we can increase the the asymmetry per decay parameter $\epsilon$,  but only if leptogenesis occurs in the strong washout regime. The reduction in  efficiency from strong washout roughly cancels the enhancement in the asymmetry per decay so that $v_R$ is again difficult to drop below $\sim 10^{12}$ GeV.

Consider a theory for neutrino masses with three gauge-singlet Weyl fermions $S_i$, that are parity even, $S_i \leftrightarrow S_i^\dag$, coupled to leptons via the interactions
\begin{align}
    \label{eq:LSi}
    {\cal L}(S_i) = S_i \,  (x^*_{ij} \, \ell_j H_L + x_{ij} \, \bar{\ell}_j H_R) +  
    \frac{1}{2} M_{S_i} \, S_i S_i + {\rm h.c.}
\end{align}
with $M_{S_i}$ real. We take 
\begin{align}
    \label{eq:largeMSi}
     M_{S_i} \gg |x_{ij}| \, v_R, \hspace{0.5in} \mbox{all } i, j 
\end{align}
so that integrating out $S_i$ leads to the dimension-5 operators of (\ref{eq:yukawaNu}). 

To see that the active neutrinos are massless at tree level, introduce a hatted basis so that the Yukawa interactions of $S_i$ can be written as 
 \begin{align}
 \label{eq:Lxi}
     {\cal L}_{x_i} = \hat{x}_i \; S_i \, (\hat{\ell}_i \, H_L +\hat{\bar{\ell}}_i \, H_R) + {\rm h.c.}, \qquad \hat{\ell_i} = x_{ij} \, \ell_j/\hat{x}_i, \qquad \hat{x}_i^2 \equiv \sum_j |x_{ij}|^2.
\end{align}
Note that the $\hat{\ell}_i$ are not orthogonal. The EFT below $M_{S_i}$ is
 \begin{align}
 \label{eq:LxiEFT}
     {\cal L}_{EFT}(S_i) \; = \; \frac{1}{2} \, \sum_i \frac{x_i^2}{M_{S_i}} \;  (\hat{\ell}_i \, H_L +\hat{\bar{\ell}}_i \, H_R)^2 + {\rm h.c.}.
\end{align}
The light neutrinos are massless at tree level  because each $S_i$ couples to only one combination of right-handed neutrinos and neutrinos, $v_R \, \hat{\nu}_{Ri} + v\, \hat{\nu}_i$, leaving the three orthogonal combinations massless. For example, in a 1-generation version of the theory the right-handed neutrino mass is $M = x^2v_R^2/M_{S}$ and the neutrino Yukawa coupling is 
$y = x^2v_R/M_{S}$, giving the correlation $y = M/v_R$. The direct and seesaw neutrino masses, defined in \ref{eq:Lnu}, are equal
 \begin{align}
 \label{eq:dir=ss}
   m_{\nu}^{\rm dir} = m_{\nu}^{\rm ss} = \frac{y^2 v^2}{M} = \frac{x^2 v^2}{M_S}. 
   \end{align}
so that the light active neutrino is massless at tree-level.

To study leptogenesis in the 3-generation theory, we abandon the hatted basis in favor of a mass basis for the right-handed neutrinos. Since the $S_i$ are integrated out, we find it convenient to first go to a non-canonical basis by rescaling $S_i \rightarrow S_i \sqrt{M_S/M_{S_i}}$, where $M_S$ is any convenient mass scale, so that the $S$ mass matrix is proportional to the unit matrix. In this basis the dimension-5 operators of the EFT are
\begin{align}
{\cal L}_\nu &= -\frac{1}{2M_S} \left( \ell_i \, (x^T x)_{ij}^* \, \ell_j \; H_L H_L  +   \bar{\ell}_i \, (x^T x)_{ij} \, \bar{\ell}_j \; H_R H_R \right)  +  \frac{1}{M_S} \,  \ell_i \, (x^\dagger x)_{ij} \, \bar{\ell}_j  \; H_L H_R + {\rm h.c.}.
\label{eq:yukawaNuS}
\end{align}
Comparing with (\ref{eq:yukawaNu}), the previously independent coupling matrices $b$ and $c$ are now correlated, with $b=x^\dagger x$, $c = x^T x$ and $M=M_S$. Thus the right-handed neutrino mass matrix and the Yukawa coupling matrix are also correlated
 \begin{align}
 \label{eq:MyinS}
   M= x^Tx \, \frac{v_R^2}{M_S}, \hspace{0.5in} y= x^\dagger x \, \frac{v_R}{M_S}. 
   \end{align}

Neutrino masses and leptogenesis thus depend on a single flavor matrix $x$. It is a general complex matrix that can be made diagonal by a bi-unitary transformation $x = U x_D V$ where $x_D$ is diagonal with entries $x_1\leq x_2 \leq x_3$. The unitary matrix $V$ can be eliminated by a transformation on the lepton doublets. Moreover, the unitary matrix $U$ contains three rotation angles and three phases, since three other phases can also be removed by transforming lepton doublet fields. A convenient form for the resulting $x$ matrix is
 \begin{align}
 \label{eq:xform}
   x = R \, e^{iA} \, x_D, \qquad A_{ij} = \epsilon_{ijk} \beta_k \, ,
   \end{align}
where the matrix $A$ is real and anti-symmetric and $R$ is a real rotation matrix, $R^T = R^{-1}$.  This parameterization is a mass basis for $\bar{\nu}$, with
 \begin{align}
 \label{eq:MyinS2}
   M_{ij}= \delta_{ij} \, M_i, \hspace{0.3in} M_i = x_i^2 \, \frac{v_R^2}{M_S}; \hspace{0.5in} y_{ij} = x_i (e^{2iA})_{ij} x_j \, \frac{v_R}{M_S}. 
   \end{align}
Remarkably, the three rotation angles of the $R$ matrix do not appear in either $M_{ij}$ or $y_{ij}$ and hence do not affect neutrino physics.

The direct and seesaw masses for the active neutrinos at tree-level are also diagonal in this basis
 \begin{align}
 \label{eq:dir=ss}
   {m_{\nu}^{\rm dir, ss}}_{ij} \; = \; \delta_{ij} \,\;  x_i^2 \, \frac{v^2}{M_S}. 
   \end{align}
Since the active neutrino mass matrix is the difference between the direct and seesaw masses, \eqref{eq:numassmatrix}, the light active neutrinos are massless at tree-level. However, this cancellation is upset by 1-loop electroweak radiative corrections.  For the case that $M_{S_i} \gg v_R \gg M_i$, the masses of the light neutrinos at leading logarithm are \cite{HHS:2023}
 \begin{align}
 \label{eq:mnuS}
   m_{\nu_i} =  \delta_i \;  x_i^2 \, \frac{v^2}{M_S} = \delta_i \; M_i \, \frac{v^2}{v_R^2},  
   \qquad \delta_i = \frac{3 g_2^2}{16 \pi^2} \left( 2\ln \frac{M_{S_i}}{v_R} + \ln \frac{v_R}{M_i} \right) \approx 0.006 \, \ln \frac{M_{S_i}^2}{v_R M_i},
   \end{align}
where we take $A=0$ for simplicity. Non-zero $A$ introduces $O(1)$ corrections.
The two logs correspond to running in the EFTs above and below $v_R$. Since the generation dependence of the loop factor $\delta_i$ is only logarithmically dependent on generation, the ratio of active neutrino masses is close to the ratio of right-handed neutrino masses, $m_{\nu_i}/ m_{\nu_j} \sim M_i/M_j$. 

To avoid tuned cancellations between terms in the matrix product $x^T x$ for the right-handed neutrino masses, the ``phases", $\beta_i$, of the antisymmetric matrix $A$ of (\ref{eq:xform}) should not be taken larger than unity.  If $\beta_i \ll 1$, the lepton asymmetry generated per $\nu_{R_1}$ decay, $\epsilon$, occurs first at order $\beta_1 \beta_2 \beta_3$ and is suppressed.  
Hence, we estimate naturalness bounds on $M_1$ and $v_R$ by studying the case of $\beta_i = {\cal O}(1)$, which gives   
 \begin{align}
 \label{eq:yijunitybeta}
   y_{ij} \simeq  x_i x_j \, \frac{v_R}{M_S} \; e^{i \phi_{ij}}, \qquad \phi_{ii}=0; \;\; \phi_{ij} = {\cal O}(1), \; i\neq j
   \end{align}
and a lepton asymmetry per $\nu_{R_1}$ decay of
 \begin{align}
 \label{eq:epsS}
   \epsilon \simeq  \frac{1}{8\pi} x_1^2 x_3^2 \, \frac{v_R^2}{M_S^2} \; \simeq  \frac{1}{8\pi} \frac{m_{\nu_3} M_1}{\delta_3 v^2} \, ,
   \end{align}
where $\nu_3$ is the heaviest of the three light neutrinos. The result for $\epsilon$ is dominated by the virtual exchange of $\nu_{R_2}$ and should be compared with (\ref{eq:eps23}): $\epsilon$ is enhanced in the singlet model by a factor $1/\delta_3$. As neutrino masses occur at 1-loop level, the Yukawa couplings required to generate the observed neutrino masses are larger than usual, enhancing $\epsilon$.  However, larger Yukawa couplings are a concern as they enhance the washout rate of the lepton asymmetry after production, and we now turn to this.

In theories where the dimension-5 operators generating the matrices $M$ and $y$ are independent, it is always possible to avoid strong washout by imposing a small contribution to the neutrino mass matrix from the seesaw exchange of $\nu_{R_1}$: $ m_{\nu_1}^{\rm ss} \lsim 10^{-3}$ eV.  However, when $M$ and $y$ matrices are correlated as in (\ref{eq:MyinS}), and the $\beta$ parameters are of order unity, the condition for weak washout becomes $m_{\nu_3}/\delta_3 \; \lsim \; 10^{-3} \mbox{eV}$. Since $m_{\nu_3} \simeq 0.05$ eV, leptogenesis is firmly in the strong washout regime, with efficiency parameter (see \eqref{eq:thermalYB})
 \begin{align}
 \label{eq:washoutS}
   \eta \simeq 10^{-2} \delta_3^{1.16}. 
   \end{align}
 Using (\ref{eq:epsS}) and (\ref{eq:washoutS}) in (\ref{eq:thermalYB}), leads to a lower bound on the mass of $\nu_{R_1}$ that is insensitive to $\delta_3$
 \begin{align}
 \label{eq:M1S}
   M_1 \gsim \frac{10^{11} \;\mbox{GeV}}{\delta_3^{0.16}} \; . 
   \end{align}
Using this result in (\ref{eq:mnuS}), the bound on $v_R$ depends on the mass of the lightest active neutrino
 \begin{align}
 \label{eq:vRS}
   v_R \gsim 1.7 \times 10^{13} \; \mbox{GeV} \; \frac{\delta_1^{0.5}}{ \delta_3^{0.08}}  \left( \frac{0.01 \, \mbox{eV}}{m_{\nu_1}} \right)^{0.5}. 
   \end{align}
In (\ref{eq:M1S}) and (\ref{eq:vRS}) the equality sign holds if $\beta_i \simeq 1$.
Therefore, in the Radiative Singlet Model, the lower bound on $v_R$ is \textit{stronger} than the naturalness limit from thermal leptogenesis with generic dimension-5 operators \eqref{eq:vRbound1}. Furthermore, $v_R$ becomes larger as $\nu_1$ is made lighter.\footnote{
In the case that $\beta_i \simeq 1$, (\ref{eq:M1S}) and (\ref{eq:vRS}) with equality signs become order of magnitude predictions for $M_1$ and $v_R$. Leptogenesis and the three light neutrino mass eigenvalues can then be used to determine $v_R$ and $x_i$. For example, with $\delta_i = 0.1$, neutrino masses of $m_{\nu_i} \simeq (0.005, 0.01, 0.05)$ eV give $x_i \simeq 0.14, 0.2, 0.45$, and $v_R \simeq 10^{13}$ GeV, where the arbitrary scale $M_S$ appearing in (\ref{eq:yukawaNuS}) - (\ref{eq:epsS}) has been set equal to $v_R$.}

Could degeneracy among the right-handed neutrinos play a role in increasing $\epsilon$ and loosening the bound of (\ref{eq:vRS}) on $v_R$? Since $\delta_i$ is only logarithmically dependent on generation, degeneracy among $\nu_R$ leads, via (\ref{eq:mnuS}), to degeneracy among the active neutrinos. Since we require
\begin{equation}
m_{\nu_2}^2-m_{\nu_1}^2=7 \times 10^{-5} \, \text{eV}^2, \quad m_{\nu_3}^2-m_{\nu_2}^2= 2.4 \times 10^{-3} \, \text{eV}^2,
\end{equation}
increasing the degeneracy of $\nu_1$ and $\nu_2$ requires increasing $m_{\nu_{1,2}}$, which is limited by the cosmological constraint on the sum of the neutrino masses,
\begin{equation}
m_{\nu_1}+m_{\nu_2}+m_{\nu_3} \leq 0.12 \, \text{eV}.
\end{equation}
This leads to a limit on the degeneracy and therefore a limit on the degeneracy factor $g(x)$ relevant for leptogenesis in Eq.  (\ref{eq:eps1}) of $g(x) \simeq m_{\nu_1}/2(m_{\nu_2}-m_{\nu_1}) \lesssim 13$. This relaxes the constraint (\ref{eq:M1S}) on $M_1$ by a factor 13 and the constraint (\ref{eq:vRS}) on $v_R$ by a factor $\sqrt{13}$. Furthermore, there is no symmetry in the Radiative Singlet Model that guarantees this degeneracy, so any weakening of (\ref{eq:M1S}) and (\ref{eq:vRS}) is accidental. Thus the Radiative Singlet Model is unable to weaken the bound on $v_R$ that results from the case of general dimension-5 neutrino mass operators.  

In the non-thermal leptogenesis scheme of Sec.~\ref{subsec:nonthermal}, the reheat temperature after decay of a degenerate gas of $\nu_{R_1}$ is below $M_1$ for $p_D \leq M_1$, as shown in  (\ref{eq:PEP}). This can lead to weak washout, even if the typical $\nu_{R_1}$ momentum at decay, $p_D$, is close to $M_1$, so that the enhancement factor of (\ref{eq:YBcomparison}) approaches 60. In addition, $\epsilon$ of (\ref{eq:epsS}) is enhanced by $1/\delta_3$, so that the lepton asymmetry is enhanced by a factor as large as $60/\delta_3$ relative to thermal leptogenesis. For $\delta_3 \sim 0.1$, the bound of (\ref{eq:vRS}) on $v_R$ is then loosened to about $4 \times 10^{10}$ GeV. While this approaches the lowest bound on $v_R$ with $\nu_R$ degeneracy, of (\ref{eq:vRbounddeg}), it requires a particular setup for reheating after inflation, and a particular $\nu_{R_1}$ decay rate. 
 
\section{Domain Wall Leptogenesis}
\label{sec:DW}

Domain walls offer another mechanism to produce right-handed neutrinos non-thermally, which can potentially enhance $Y_1$ above the thermal value and  relax the naturalness bound on $v_R$. In this section we show that non-thermal leptogenesis produced by the production of $\nu_{R_1}$ from the decay of domain walls requires $v_R \gtrsim 2 \times 10^{12}$ GeV and thus   \textit{cannot} beat the naturalness bound of $v_R \gtrsim 10^{12} \, \rm GeV$ \eqref{eq:vRbound1}.

Domain walls are interesting to LR models since they are naturally inherent to the theory; the breaking of the discrete symmetry associated with the LR symmetry generates domain walls at the scale $v_R$ \cite{Kibble:1976sj,everett1982left,Kibble:1982dd}. Oftentimes, the default assumption is that the reheat temperature of the universe after inflation is below the scale $v_R$ so that the domain walls are effectively diluted away and can never come to dominate the universe - a generic problem of domain walls known as the `domain wall problem'. Nevertheless, this assumption need not be true and it is possible that the reheat temperature after inflation was above the scale $v_R$ so that domain walls formed in our early universe after inflation. Compatibility with our present-day universe then requires that the walls decay before dominating the universe\footnote{Domain walls experience repulsive gravitational accelerations of order $G \sigma$ \cite{Ipser:1983db,Sikivie:2006ni}. This repulsive pressure, $\sim G \sigma^2$ always dominates over any vacuum pressure difference on the walls if the vacuum pressure is weak enough to allow the walls to dominate the energy density of the universe in the first place.},
which can result from a small parity breaking term in the theory or, depending on the GUT completion, from becoming attached to cosmic strings, causing the wall-bounded string system to decay via gravitational waves \cite{everett1982left,Martin:1996ea,Dunsky:2021tih}.  

We now proceed with a few simple arguments to show that leptogenesis via non-thermal $\nu_{R}$ production from domain wall annihilation does not allow values of $v_R$ below $10^{12}$ GeV, the lower bound from thermal leptogenesis.

\begin{figure}[tb]
    \centering
        \includegraphics[width=.45\textwidth]{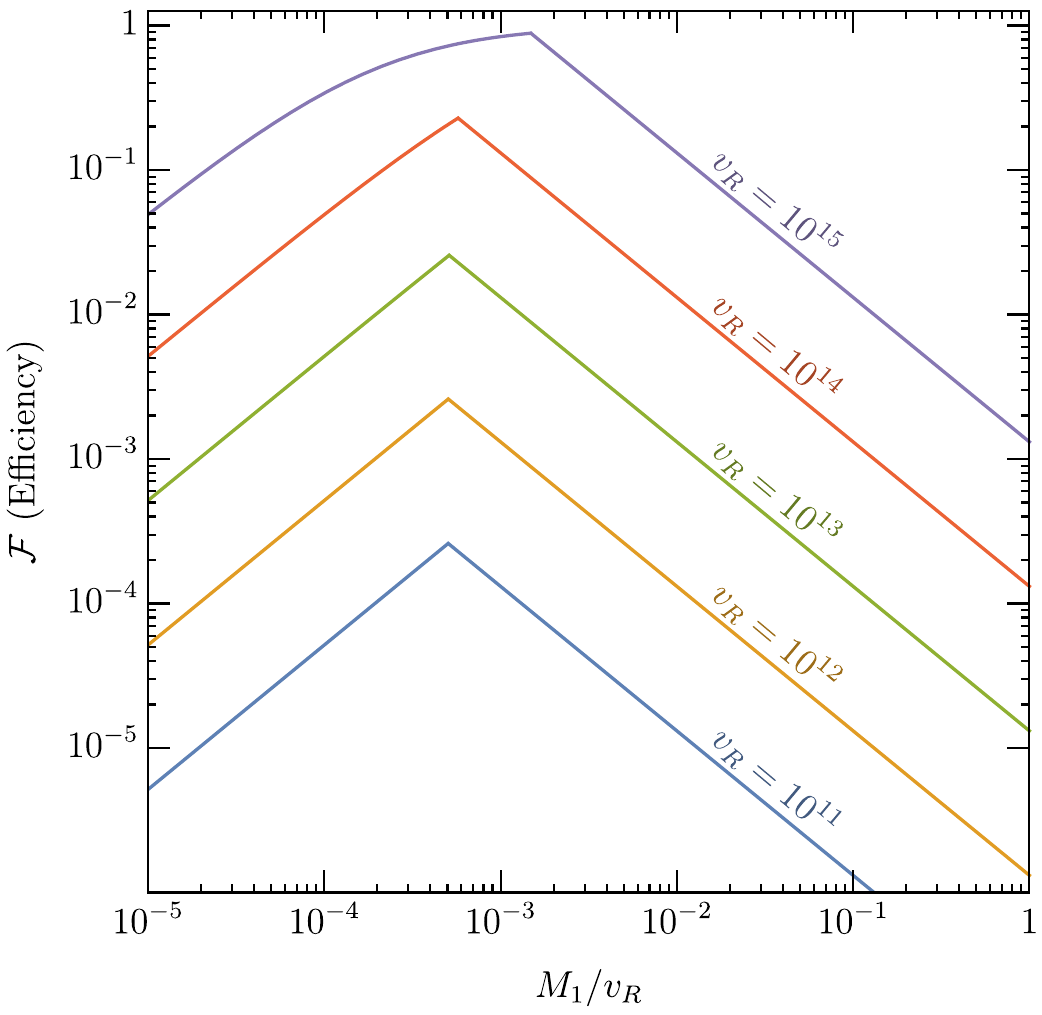} 
        \hfill
        \includegraphics[width=.45\textwidth]{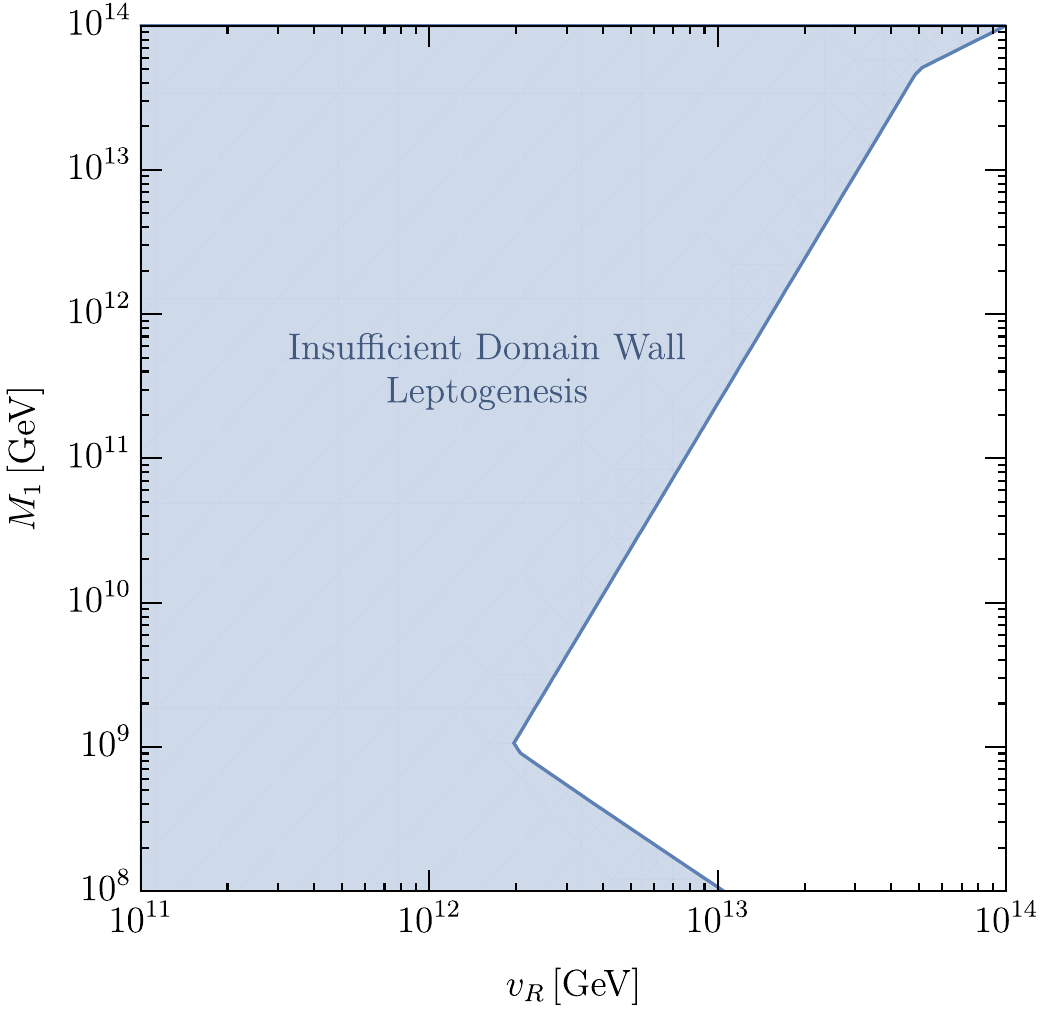}
    \caption{\textbf{Domain Wall Leptogenesis.} \textbf{Left}: Efficiency parameter $\mathcal{F}$ (fraction of $h$ that decay to $\nu_{R_1}$) as a function of $M_1/v_R$. For small $M_1/v_R$, $\mathcal{F}$ is set by the branching ratio of $h \rightarrow \nu_{R_1} l_1$, corresponding to the first argument of the $\textit{Min}$ function in Eq. \eqref{eq:efficiency}; the branching ratio increases linearly with $M_1$ and hence $\mathcal{F}$ rises linearly. For larger $M_1/v_R$, $\mathcal{F}$ is set by the fraction of decays  that occur when $h$ can kinematically decay to $\nu_{R_1}$, which occurs at times much earlier than the lifetime of $h$, corresponding to the second argument of the $\textit{Min}$ function in Eq. \eqref{eq:efficiency}; the kinematically allowed fraction of decays decreases linearly with $M_1$ and hence $\mathcal{F}$ decreases linearly. \textbf{Right}: Domain wall leptogenesis can generate the observed baryon asymmetry, using the efficiency $\mathcal{F}$ of the left panel, in the unshaded region of the $M_1 - v_R$ plane, requiring $v_R \gtrsim 10^{12}$ GeV.}
    \label{fig:DWLeptogenesis}
\end{figure}

Without loss of generality, let $\nu_{R_1}$ be the right-handed neutrino whose decays are responsible for leptogenesis. The non-thermal yield of $\nu_{R_1}$ from wall annihilation can be written as
\begin{align}
    \label{eq:YN1_DW}
    Y_{\nu_{R_1}} = \mathcal{F} \; \frac{\rho_{\rm DW}(t_\Gamma)}{M_1 s(t_\Gamma)}
\end{align}
where $\rho_{\rm DW}(t_\Gamma)$ is the energy density of walls at the wall annihilation time, $t_\Gamma$, $s(t_\Gamma)$ is the entropy density at $t_\Gamma$. The efficiency factor $\eta \leq 1$ parameterizes how effectively the wall energy is transmitted to right-handed neutrinos; it is at most unity when all the energy density of the walls is transmuted to non-relativistic $\nu_{R_1}$. In general, $\mathcal{F}$ is the fraction of $h$ that end up decaying to $\nu_{R_1}$. As discussed in Appendix \ref{app:DW}, $\mathcal{F}$ depends on the product of the branching ratio of $h \rightarrow \nu_{R_1} l_1$ and the fraction of decays that occur in the time when $h$ can kinematically decay into $\nu_{R_1}$. We estimate $\mathcal{F}$ in Appendix \ref{app:DW} and find it to be much less than $10^{-3}$ for $v_R < 10^{12}$ GeV. The left panel of Fig.~\ref{fig:DWLeptogenesis} shows the dependence of $\mathcal{F}$ on $M_1/v_R$ for a variety of $v_R$.

The baryon yield eventually generated by the $\nu_{R_1}$ originating from domain walls is
\begin{align}
    \label{eq:YBMax}
    Y_B = \frac{28}{79} Y_{\nu_{R_1}} \epsilon_1  \leq \frac{28}{79} Y_{\nu_{R_1}} \epsilon_{\rm Max} \, ,
\end{align}
where $\epsilon_{\rm Max} \approx \frac{3}{16\pi}{M_1 \Delta m_\nu}/v_L^2$ is the maximum asymmetry per decay \cite{Hamaguchi:2001gw,Davidson:2002qv,Davidson:2008bu} when the right-handed neutrinos are not highly degenerate \cite{Hambye:2003rt}. Inserting \eqref{eq:YN1_DW} into \eqref{eq:YBMax} implies the maximum baryon asymmetry is independent of the right-handed neutrino mass, $M_1$, such that 
\begin{align}
    Y_B \lesssim \mathcal{F} \frac{0.05}{g_{*s}}  \frac{\Delta m_\nu}{v_L^2} \frac{\rho_{\rm DW}(t_\Gamma)}{T_\Gamma^3} \,.
\end{align}
The value for $\rho_{\rm DW}/{T_\Gamma}^3$ depends on the cosmological state of the walls at decay,
\begin{align}  
    \label{eq:DWFormFactor}
    \frac{\rho_{\rm DW}(t_\Gamma)}{T_\Gamma^3} \approx v_R
    \times \begin{dcases}
        \left(\frac{v_R}{M_{\rm Pl}} \right)^{\frac{1}{3}} \lambda^{\frac{1}{2}} & \text{(Wall formation, Kibble-Zurek)}
        \\
        \left(\frac{v_R}{M_{\rm Pl}} \right)^{\frac{1}{2}} (2 C \beta \lambda^{\frac{1}{2}})^{\frac{1}{2}} & \text{(Friction Regime)}
        \\
        \left(\frac{v_R}{M_{\rm Pl}} \right)^{\frac{1}{2}} (8 C^3 \lambda^\frac{1}{2})^{\frac{1}{2}} \left(\frac{t}{t_{\rm dom}} \right)^\frac{1}{2} & \text{(Scaling Regime)}
    \end{dcases}
\end{align}
Above, we have written the wall tension as $\sigma = \lambda^{1/2} v_R^3$, where $\lambda$ is the Higgs quartic coupling. The variable $\beta$ parameterizes the number of particles scattering with the wall and is always above unity due to the interactions of the wall scalar field with the thermal bath. Last, $C^2 = 8\pi^3 g_*/90$ and $t_{\rm dom} \equiv M_{\rm Pl}^2/\sigma$ is the time at which the walls dominate the density of the universe. 

We see that Eq. \eqref{eq:DWFormFactor} is maximized when the walls decay soon after formation due to the enhanced Kibble-Zurek initial abundance. However, even with this maximum possible $Y_B$, there remains a strong lower bound on $v_R$ in order to achieve a sufficient baryon asymmetry; namely, to match the observed baryon abundance, $Y_B^{(\rm obs)} \simeq 8 \times 10^{-11}$, $v_R$ must be greater than $\sim 6 \times 10^{10}$ GeV assuming an unrealistic maximal efficiency $\mathcal{F} = 1$. In reality, $\mathcal{F}$ decreases dramatically for lower $v_R$ since the efficiency is maximized when the mass of $\nu_{R_1}$ matches that of decaying Higgs field so it as non-relativistic as possible. Consequently, when incorporating the reduced efficiency of $Y_{\nu_{R_1}}$ as shown in the left panel of Fig.~\ref{fig:DWLeptogenesis}, the lower limit on $v_R$ increases to $\sim 2 \times 10^{12}$ GeV, as shown by the solid blue region in the right panel of Fig.~\ref{fig:DWLeptogenesis}. This lower limit for $v_R$ is comparable to the naturalness limit from thermal leptogenesis.

\section{Summary}
\label{sec:summary}

In theories where the electroweak sector includes $SU(2)_L \times SU(2)_R$ with minimal Higgs doublets $H_L(2,1)$ and $H_R(1,2)$, an approximate parity symmetry can solve the strong CP problem over a wide range of $\vev{H_R} = v_R$. When parity is exact, spontaneous parity breaking occurs via the radiative Higgs Parity mechanism, and the observed values of $m_t$ and $\alpha_s$ require $v_R > \, 10^9 \, \GeV$. Neutrino masses, in both LR and Mirror versions of the theory, arise from an effective theory with dimension-5 operators restricted by parity. This correlates the light neutrino masses with the right-handed neutrino masses and neutrino Yukawa couplings. Requiring that the lightness of the observed neutrinos does not involve fine-tuning, we have shown that thermal leptogenesis requires $v_R \gsim 10^{12}$ GeV, provided there is no degeneracy among the right-handed neutrinos.\footnote{The theory does have fine-tuning in the SM Higgs boson mass. However, the electroweak hierarchy may originate from some mechanism that is not available in the neutrino sector; for example, environmental selection in a multiverse.} 

This bound on $v_R$ leads to an indirect probe of leptogenesis in this theory. 
The large values of $v_R$ required for leptogenesis are consistent with the observed Higgs mass only for low values of $m_t$ and/or high values of $\alpha_s$, as shown in the right panel of Figure \ref{fig:vpPrediction}. More accurate determinations of $m_t$ and $\alpha_s$ will greatly reduce the uncertainties in the prediction for $v_R$, as shown in Fig.~\ref{fig:vpPredictionFuture} for the LR model and Fig.~\ref{fig:vpPredictionFuturevp} for the Mirror model. If $v_R$ is determined to be below $10^{12}$ GeV, thermal leptogenesis would be generically excluded. On the other hand, a determination of $v_R$ above $10^{12}$ GeV would lend support to thermal leptogenesis. Furthermore, a neutron electric dipole moment becomes more likely as $v_R$ increases, generated by the operator of (\ref{eq:dim6}). Over the coming decade, or beyond, a determination of a large value for $v_R$ and a discovery of an electric dipole moment of the neutron would together provide significant indirect evidence that the cosmological baryon asymmetry was created via thermal leptogenesis in this theory.

The LR theory may be embedded into $SO(10)$ unified theories~\cite{Hall:2019qwx}.
The preferred $v_R$ and unification scale from precise gauge coupling unification depend on the mass spectrum of heavy gauge bosons which is determined by the representations of the $SO(10)$-breaking Higgses. 
If $SO(10)$ is broken only by a ${\bf 45}$ Higgs, precise gauge coupling unification prefers $v_R$ and the  unification scale to be around $10^{11}$~GeV and $10^{17}$~GeV, respectively. If a ${\bf 54}$ Higgs also obtains a vev, which helps stabilize the desired vacuum~\cite{Babu:1984mz}, the preferred $v_R$ increases to $10^{12-13}$ GeV and is consistent with the requirement from successful leptogenesis. At the same time, the preferred unification scale decreases so that future observations of proton decay are more likely.

If future experiments find values of $m_t$ and $\alpha_s$ such that $v_R$ is determined to be less than $10^{12}$ GeV, then there are four possibilities for retaining successful leptogenesis:
\begin{itemize}
\item Adding extra scalars to the theory below $v_R$. A mixed quartic coupling with the SM Higgs affects the running of the SM quartic coupling, increasing $v_R$.  This extra light scalar increases the fine-tuning of the theory.\footnote{We note that the Higgs Parity theory has the same amount of fine-tuning in the Higgs potential as in the SM.  While it does not address the hierarchy problem, it does not make it worse.} 
\item Adding soft breaking of parity in the potential for $H_L$ and $H_R$. This allows a tree-level vacuum with a large hierarchy of vevs, so that the SM quartic is unconstrained. However, the soft breaking is large, typically with a scale of order $v_R$, and its origin requires a complication of the theory. 
\item Leptogenesis may be non-thermal, for example resulting from the decays of a degenerate gas of $\nu_{R_1}$ after inflation. However, this still requires $v_R > 10^{11}$ GeV. 
\item Degeneracy can be imposed in the right-handed neutrino spectrum.  Degeneracy at the level of $10^{-6}$ may naturally result from approximate symmetries and can lead to reductions in $M_1$ to below $10^4$ GeV and $v_R$ to near $10^{9}$ GeV. This would be consistent with values of $m_t(\alpha_s)$ that are 3$\sigma$ above (below) the current central values.
\end{itemize}
The first two options significantly weaken the simplicity of the theory; only $\nu_R$ degeneracy easily allows much lower values of $v_R$.

If parity is softly broken, even by a small amount, parity may break after inflation producing a domain wall network. We find that leptogenesis during collisions of these domain walls does not avoid the bound of $v_R \gsim 10^{12}$ GeV. 

The dimension-5 neutrino mass operators can result from the exchange of fermions that are gauge singlets. 
This Radiative Singlet Model is particularly simple, and correlates parameters so that the direct and seesaw contributions to the light neutrino masses cancel. 
Neutrino masses are radiative, and hence require larger Yukawa couplings.  While this enhances the creation of the lepton asymmetry, it forces thermal leptogenesis into the strong washout domain; the latter dominates, so that the natural bound on $v_R$ is strengthened.  In this model, the prospect for discovery of a neutron electric dipole moment is excellent. In the special case that the Radiative Singlet Model yields a lepton asymmetry via non-thermal leptogenesis, strong washout can be avoided. The bound on $v_R$ in the third bullet above can then be weakened to $v_R > 4 \times 10^{10}$ GeV. 

\begin{figure}[tb]
     \centering
     \includegraphics[width=.8\textwidth]{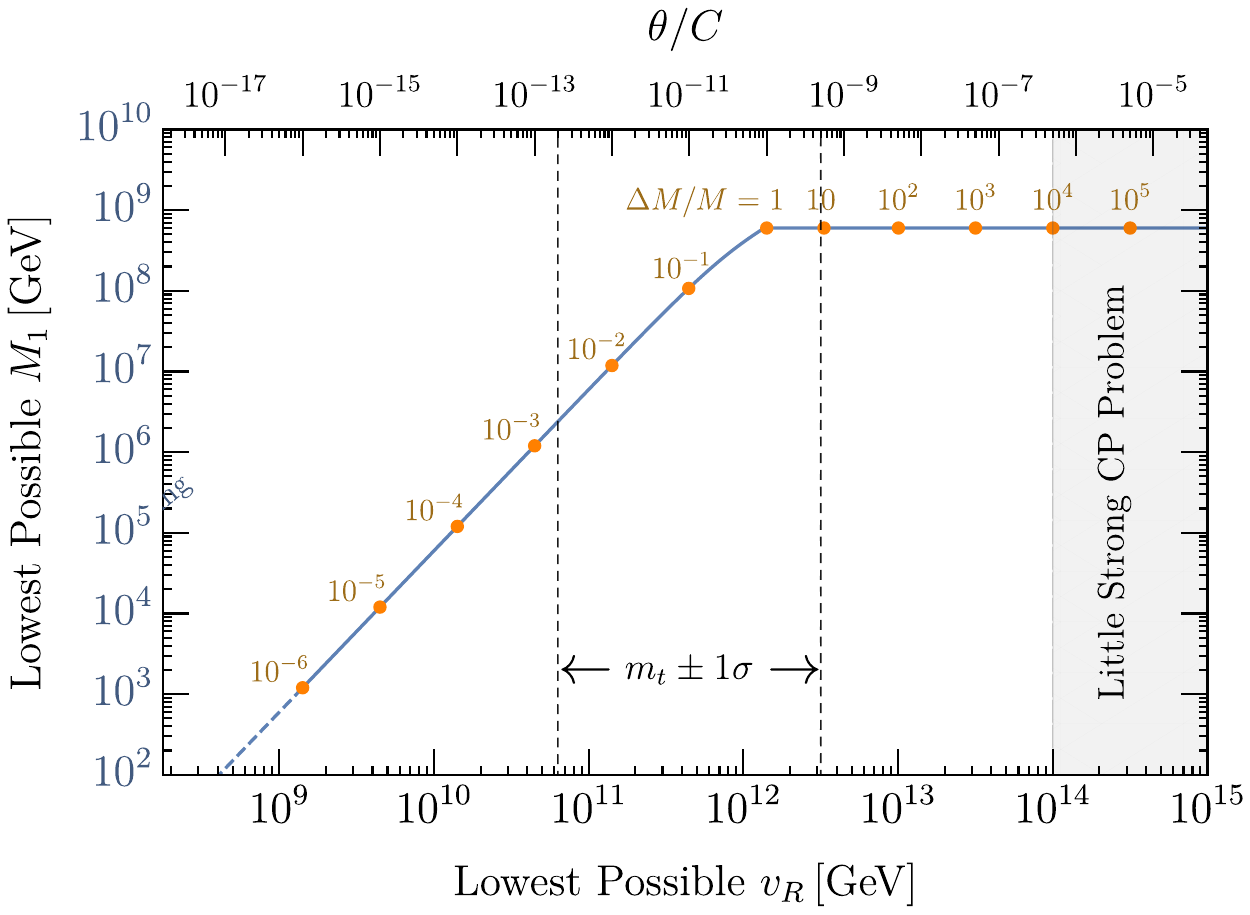} 
     \caption{Lower bounds on $v_R$ (horizontal axis) and $M_1$ (vertical axis) from thermal leptogenesis in theories with exact parity and minimal Higgs scalars, as a function of the degeneracy $\Delta M/M$ of the right-handed neutrinos involved in leptogenesis. In the left of the figure, in the region where there is a degree of degeneracy between $\nu_{R_1}$ and $\nu_{R_2}$, we take $M_3 \sim M_1$; to the right of the figure, where $M_2 \gg M_1$, we take $M_3 \lesssim M_2$. Other choices for $M_3$ give stronger lower bounds on $v_R$. The curve traces out the intersection of the blue and orange lines in Figs.~\ref{fig:naturalnessPlot} and \ref{fig:naturalnessPlotdegen} in the non-degenerate and degenerate regime, respectively.
     The contour is dashed where the high degree of right-handed neutrino degeneracy can only be obtained by fine-tuning. 
     The dashed vertical lines show the predicted range for $v_R$ for $m_t = (172.56 \pm 0.48) \; \GeV $.
     As $v_R$ is increased, the dimension-6 operator of (\ref{eq:dim6}) gives a contribution to $\theta$ proportional to $v_R^2$, as indicated by the top axis. Increasing $v_R$ from $10^{12} \, \GeV$ to $10^{14} \, \GeV $, current experiments searching for the neutron electric dipole moment are progressively more likely to see a signal. Increasing $v_R$ above $10^{14}$ GeV, the strong CP problem reappears, as show by the gray shading.
     }
     \label{fig:naturalnessPlotSummary}
 \end{figure}
 
Fig.~\ref{fig:naturalnessPlotSummary} summarizes our results for thermal leptogenesis in the minimal Higgs Parity theories that solve the strong CP problem via an exact parity, with neutrino masses arising from the dimension-5 effective theory of (\ref{eq:yukawaNu}). The blue curve shows the lowest possible value of $v_R$ and $M_1$ as a function of $\Delta M/M = (M_2 - M_1)/M_1$ which, when small, is a measure of the degeneracy between the decaying and virtual $\nu_R$. The orange dots show representative values of $\Delta M/M$ so that degeneracy increases from right to left. The horizontal (diagonal) segment represents the non-degenerate (degenerate) region and corresponds to the intersection of the blue and orange lines of Fig.~\ref{fig:naturalnessPlot} (\ref{fig:naturalnessPlotdegen}), which represent the lowest possible values of $v_R$ and $M_1$ for a given $\Delta M/M$.

Fig.~\ref{fig:naturalnessPlotSummary} demonstrates that the lowest possible $v_R$ and $M_1$ require the most degenerate $\nu_R$. However, 
flavor breaking in the charged leptons limits the natural degeneracy of right-handed neutrinos to the level of $10^{-6}$ as shown by the dashed blue curve, which corresponds to $(v_R, M_1)$ below $(10^9 \, {\rm GeV},  10^3 \, {\rm GeV})$. On the other hand, for large $v_R$, the discovery of a neutron electric dipole moment becomes more likely as it can be generated by the operator of (\ref{eq:dim6}). A signal is likely if $v_R$ is of order $10^{12}$ GeV, the lower bound from leptogenesis without degeneracy, and is strongly expected for $v_R$ above $10^{13}$ GeV. As $v_R$ increases above $10^{14}$ GeV, the strong CP problem begins to re-emerge, as shown by the gray shading. When parity is exact, the scale of its spontaneously breaking, $v_R$, can be computed from SM parameters. For $\alpha_s(M_Z) = 0.1179$ and the 1$\sigma$ range of $m_t = (172.56 \pm 0.48) \GeV$, the predicted range for $v_R$  is shown by the vertical dotted lines. Of course, including uncertainties in $\alpha_s$ and more than $1 \sigma$ in $m_t$ the allowed range is much wider, but future measurements offer the prospect of a much more accurate prediction, with important implications for leptogenesis and neutrino masses.

\begin{acknowledgments}
We thank the CERN Theory Division where part of this work was performed. DD is supported by the James Arthur Postdoctoral Fellowship. The work of JCCM and LH was supported by the Director, Office of Science, Office of High Energy Physics of the U.S. Department of Energy under the Contract No. DE-AC02-05CH11231 and by the NSF grants PHY-1915314 and PHY-2210390.
The work of KH was partly supported by Grant-in-Aid for Scientific Research from the Ministry of Education, Culture, Sports, Science, and Technology (MEXT), Japan (20H01895) and by World Premier International Research Center Initiative (WPI), MEXT, Japan (Kavli IPMU).
\end{acknowledgments}

\appendix 
\section{Naturalness Limit on $\nu_{R}$ Degeneracy}
\label{app:degeneracy}
In a natural theory, the degeneracy between $\nu_{R_1}$ and $\nu_{R_2}$ should be greater than any quantum corrections that generate a mass difference between them. In LR models, the operators which generate the masses for the charged leptons also generate unavoidable corrections to $M_1$ and $M_2$; that is, they break the symmetry that made $\nu_{R_1}$ and $\nu_{R_2}$ degenerate. For example, note that the operators that describe the neutrinos Eq. \eqref{eq:yukawaNu}, and the charged fermions,
\begin{align}
\label{eq:chargedFermions}
\mathcal{L}_{e,u,d} =\frac{c_{ij}^u}{M}q_i\bar{q}_j H_L H_R + \frac{c_{ij}^d}{M}q_i\bar{q}_j H_L^\dagger H_R^\dagger + \frac{c_{ij}^e}{M}l_i\bar{l}_j H_L^\dagger H_R^\dagger + \text{h.c.} \, ,
\end{align}
are necessarily described by an effective field theory; the dimension-5 operators given in Eq. \eqref{eq:chargedFermions} \text{must} be generated by other physics. There are two possible ways to generate such operators: by integrating out a heavy scalar field or by integrating out a heavy fermion field.  

\begin{figure}[tb]
    \centering
        \includegraphics[width=.45\textwidth]{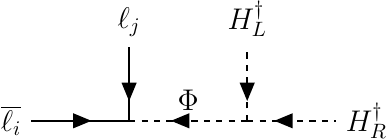} 
        \hfill
        \includegraphics[width=.45\textwidth]{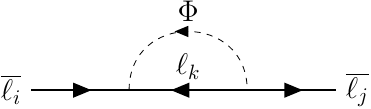}
    \caption{Left: Feynman diagram showing how the charged lepton Yukawa couplings are generated in a UV completion involving a bifundamental scalar. Right: The same interaction also necessarily generates a mass splitting for the right-handed neutrinos, limiting the smallest natural degeneracy between $\nu_R$'s to be of order $\Delta M/M \sim y_\tau^2/8 \pi \sim 10^{-6}$. 
    }
    \label{fig:scalarFeynman}
\end{figure}

First, consider the case where the charged lepton Yukawas arise from integrating out a bifundamental scalar in the UV completion
\begin{align}
\label{eq:Lscalar}
{\mathcal L } =    - m_{\Phi}^2 |\Phi|^2 + (x_{ij} \Phi \ell_i \bar{\ell}_j  - A \Phi^\dag H_L^\dag H_R^\dag + {\rm h.c.}) \, ,
\end{align}
as shown in the left panel of Fig.~\ref{fig:scalarFeynman}. This generates the charged lepton Yukawa coupling $y^e_{ij} = A x_{ij} v_R/m_{\Phi}^2 $.
It also also generates non-canonically normalized kinetic terms for the $\nu_R$,
\begin{align}
    \label{eq:noncanicalKinetic}
     {\cal L} = & ( 1 + \delta Z_{11}) \; \bar{\nu}_1^\dag \bar{\sigma}\partial \bar{\nu}_1 +( 1 + \delta Z_{22}) \; \bar{\nu}_2^\dag \bar{\sigma}\partial \bar{\nu}_2 + \left( \delta Z_{12} \, \bar{\nu}_1^\dag \bar{\sigma}\partial \bar{\nu}_2 + {\rm h.c.} \right),
\end{align}
where $\delta Z_{ij} \simeq x_{ki}x^*_{kj}/8\pi^2$,
as shown by the diagram on the right panel of Fig.~\ref{fig:scalarFeynman}. This wave function renormalization induces a mass splitting of \cite{Dunsky:2020dhn} 
\begin{align}
    \label{eq:degeneracyScalar}
    \frac{\Delta M}{M}  \simeq \sqrt{\left( \delta Z_{11} - \delta Z_{22} \right)^2 + \left( \delta Z_{12} + \delta Z_{12}^* \right)^2 } \gtrsim \frac{y_\tau^2}{8\pi^2} \simeq 10^{-6} \, ,
\end{align}
where $y_\tau \simeq 10^{-2}$ is the $\tau$ Yukawa coupling.

\begin{figure}[tb]
    \centering
        \includegraphics[width=.45\textwidth]{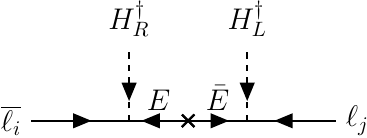} 
        \hfill
        \includegraphics[width=.45\textwidth]{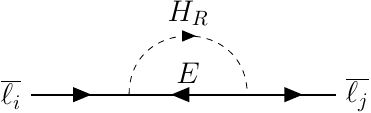}
    \caption{Same as Fig.~\ref{fig:scalarFeynman} but when the charged lepton Yukawa couplings are generated by heavy fermions $E,\bar{E}$. The naturalness limit on the radiative mass splitting is also $\Delta M/M \sim y_\tau^2/8\pi \sim 10^{-6}$.}
    \label{fig:fermionFeynman}
\end{figure}

A similar constraint occurs if instead the charged lepton Yukawas arise from integrating out heavy fermions, $E, \bar{E}$, in the UV completion 
\begin{align}
\label{eq:Lfermion}
 {\mathcal L} = z_{ia}^e \ell_i \bar{E}_a H^\dag_L + (z^e_{ia})^* \bar{\ell}_i E_a H^\dag_R + M_{E,a} E_a \bar{E}_a \; + \text{h.c.} ,
\end{align}
as shown in the left panel of Fig.~\ref{fig:fermionFeynman}. 
This generates the charged lepton Yukawa coupling $y_{ij}^e = z^e_{ia} (z_{aj}^{e})^* v_R/M_{E,a}$ if $M_{E,a} > z_e v_R$ or $y_{ij}^e = z_e$ if $M_{E,a} < z_e v_R$.
Again, the same operators in \eqref{eq:Lfermion} generate non-canonically normalized kinetic terms as in Eq. \eqref{eq:noncanicalKinetic} now with $\delta Z_{ij} \simeq z_{ia}^e(z_{ja})^*/8\pi^2$.  Wavefunction renormalization thus gives the same limit on $\Delta M/M$ as Eq. \eqref{eq:degeneracyScalar}.

\begin{figure}[tb]
    \centering
        \includegraphics[width=.45\textwidth]{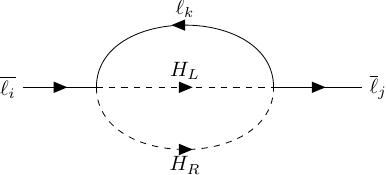} 
     \caption{ In the effective field theory, radiative corrections generate dimension-5 operators for neutrino masses from the dimension-5 interaction that leads to the charged lepton Yukawa couplings.}
    \label{fig:radiative_Diagram}
\end{figure}

Last, even without reference to any UV completion, in the LR effective field theory of Eq. \eqref{eq:chargedFermions}, the dimension-5 operators for the charged lepton masses necessarily generate a mass splitting for $\nu_{R,i}$. This is demonstrated by the diagram of Fig.~\ref{fig:radiative_Diagram} which generates a wave function renormalization parametrically similar  to that generated in the two aforementioned UV completions,
\begin{align}
    \label{eq:degeneracyEFT}
    \frac{|\Delta M|}{M} \sim \frac{1}{(16\pi^2)^2} y_\tau^2 \left(\frac{\Lambda}{v_R}\right)^2 \sim 5 \times 10^{-9} \left(\frac{\Lambda}{v_R}\right)^2
\end{align}
which is roughly one-loop factor smaller than Eq. \eqref{eq:degeneracyScalar}, though it can be larger for $\Lambda \gg v_R$.

\section{Efficiency of Domain Wall Leptogenesis}
\label{app:DW}
In this section, we elaborate on the production of $\nu_R$ from the annihilation of domain walls and estimate the efficiency parameter, $\eta$, of Eq. \eqref{eq:YN1_DW}, showing that it is highly suppressed for $v_R \lesssim 10^{12}$ GeV and becomes $\mathcal{O}(1)$ only for $v_R \gg 10^{12}$ GeV. Consequently, domain wall leptogenesis is not a promising way to lower the naturalness bound on $v_R$ below $10^{12}$ GeV.

\begin{figure}[tb]
    \centering
    \includegraphics[width=.95\textwidth]{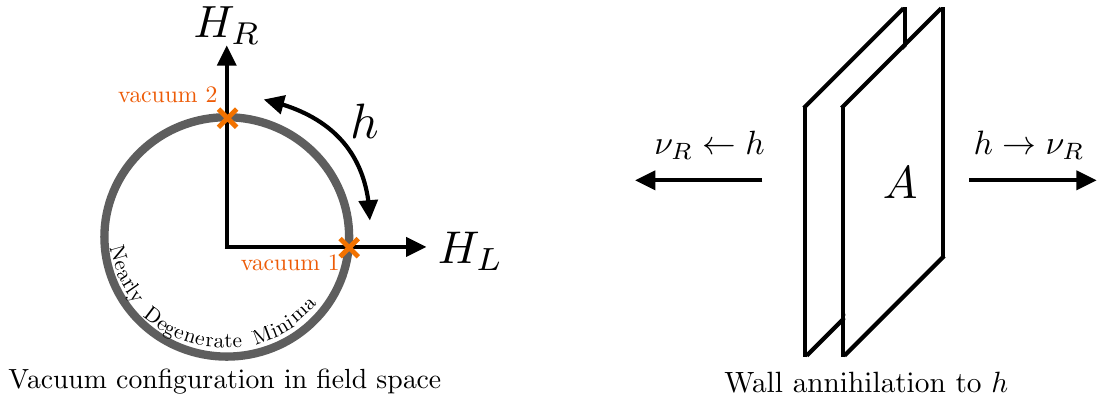} 
    \caption{\textbf{Left:} Visualization of vacuum configuration in field space for $H_L$ and $H_R$. Because the Coleman-Weinberg potential and the $Z_2$-breaking piece are small (see Eq. \eqref{eq:generalPotential}), the potential is nearly $SU(4)$ symmetric with the the angular excitations $h$ being nearly massless. The curvature of the Coleman-Weinberg potential generates a small mass for the $H_L$ vacuum while the $Z_2$-breaking piece ensures that this minimum is the absolute minimum of the potential, causing the domain walls to annihilate. \textbf{Right:} Simplified illustration of leptogenesis from domain wall annihilation. If there exists a small parity breaking term in the theory, a pressure difference is generated between the walls, eventually causing them to collide and annihilate. After annihilation, excitations of the angular field $h$  emanate from the surface of contact of the walls, representing excitation of Higgs quanta. These Higgs quanta initially have large field values and can thus kinematically decay to $\nu_R$, potentially generating a large non-thermal abundance of $\nu_R$ whose own decays eventually generate a lepton asymmetry.}
    \label{fig:DWAnnihilation}
\end{figure}

Consider two domain walls immediately before coming into contact and annihilating into Higgs particles. Sufficiently close to the walls, the surfaces are roughly plane parallel patches of area $A$ as shown in Fig.~\ref{fig:DWAnnihilation}. The energy in each wall patch is
\begin{align}
    E_w &\simeq \gamma(v_w) \sigma A, \quad \gamma(v_w) = \frac{1}{\sqrt{1-v_w^2}} \ .
\end{align}
The Lorentz factor of the walls is roughly unity since friction between the wall surface and the background plasma impedes wall motion even with the false vacuum accelerating the walls together.

After wall annihilation, Higgs particles are produced. We have performed 1D simulations of planar walls annihilating and find that after a time $t$, the Higgs field approximately spreads out uniformly in space out to a distance $L \approx t$. The average energy density of the Higgs field $h$ (the angular excitation between $H_L$ and $H_R$ as shown in Fig.~\ref{fig:DWAnnihilation})     within a distance $L  \ll t_{\Gamma}$ from the point of annihilation is then
\begin{align}
    \rho_h \simeq \frac{E_w}{A \times L}  = \frac{\gamma(v_w) \sigma}{L} \approx \lambda h^4.
\end{align}
Since the wall tension of the Higgs field is approximately $\sigma \simeq \sqrt{\lambda}v_R^3$ and the Higgs fields propagates relativistically, the value of the Higgs field a time $t \simeq L$ after annihilation is
\begin{align}
    h \approx \left(\frac{v_R^3}{\lambda^{1/2} t} \right)^{1/4} \,. 
\end{align}
We see that at early times after wall annihilation, the Higgs field is largest. Kinematic production of $\nu_{R_1}$ requires that the effective Higgs mass, $m_{{\rm eff},h} \approx \sqrt{\lambda}h$, is greater than $M_1$. This occurs at a time less than
\begin{align}
    t_{\rm max} = \lambda^{3/2} \frac{v_R^3}{M_1^4} \, .
\end{align}
When $t_{\rm max} < \Gamma_{h \rightarrow \nu_{R_1} l_1}^{-1}$, the production of $\nu_{R_1}$ is suppressed since only a small fraction of decays occur while $h$ is large enough to permit the reaction $h \rightarrow \nu_{R_1} l_1$. Quantitatively, the fraction of $\nu_{R_1}$ ultimately produced -- equivalently, the efficiency factor $\mathcal{F}$ -- is 
\begin{align}
    \mathcal{F} = \text{Br}(h \rightarrow \nu_{R_1} l_1)\times {\text{Min}}\left\{1, \,\Gamma_h t_{\rm max} \right\} \simeq \frac{y_1^2}{y_b^2 + y_1^2} \times {\text{Min}}\left\{1, \, \frac{\lambda}{8\pi} (y_1^2+ y_b^2) \frac{v_R^2}{M_1^2} \right\}  \,.
    \label{eq:efficiency}
\end{align}
Here, $\Gamma_h \simeq (y_1^2+ y_b^2)/(8\pi) \sqrt{\lambda} h (h/v_R)$ is the total decay rate of the Higgs particles emanating from the wall, $y_b$ the bottom quark Yukawa coupling, and $h/v_R$ is the time dilation factor of the Higgs particles escaping from the wall which have momenta $k \approx \sqrt{\lambda} v_R$ and mass $m_{{\rm eff},h} \approx \sqrt{\lambda}h$. Note that since the effective top quark mass $m_{{\rm eff},t} \approx y_t h$ is greater than $m_{{\rm eff},h}$, $h$ cannot decay to two top quarks. Instead, the dominant $h$ decay out of $\nu_{R_1} l_1$ is to two bottom quarks. For $y_1 < y_b$, the efficiency is suppressed by the small branching ratio into $\nu_{R_1}$ compared to bottom quarks. We take the renormalized value of $y_b$ above $v_R > 10^{10}$ GeV to be around $0.008$.

We show the value of $\mathcal{F}$ as a function of $M_1/v_R$ in the left panel of Fig.~\ref{fig:DWLeptogenesis}. Here, we take $y_1^2 = M_1 \tilde{m}/v^2$ where $\tilde{m} = 0.1$ eV is the largest natural value for $\tilde{m}$ without needing fine-tuned cancellations between the direct and seesaw mass contributions of $\nu_1$. Note that since domain wall leptogenesis is non-thermal, it is possible to take $y_1 > y_b$ without causing washout which can occur for such large $\tilde{m}$. We also take $\lambda = 0.1$ which is an optimistically large value of the Higgs quartic near the scale $v_R$. The right panel of Fig.~\ref{fig:DWLeptogenesis} shows the region where the baryon asymmetry given by Eq. \eqref{eq:YN1_DW},  and using the efficiency of Eq. \eqref{eq:efficiency}, is unable to match the observed baryon $Y_B \simeq 8 \times 10^{-11}$. We see that domain wall leptogenesis requires $v_R \gtrsim 5 \times 10^{12}$ GeV, comparable to the bound from natural thermal leptogenesis.

\section{Model-dependence of Parity Breaking Scale}
\label{sec:model dependence}
In this appendix, we discuss the model dependence of the Parity breaking scale arising from the threshold correction at the parity breaking scale and the running of the quartic coupling. We discuss Models A and D of Ref.~\cite{Hall:2018let}, which have no additional stable charged particles and are consistent with cosmological evolution with high enough reheat temperature to achieve thermal leptogenesis. The fermion contents of Models A and D are shown in Tables~\ref{tab:chargesA} and \ref{tab:chargesD}. Model A with large Dirac masses, $M_{U_i} U_i \bar{U}_i + M_{D_i} D_i \bar{D}_i + M_{E_i}E_i \bar{E}_i$, gives the LR theory. Integrating out the heavy fermions generates dimension-5 operators leading to SM quark and charged lepton masses after electroweak symmetry breaking. Model $D$ is the Mirror model.

\begin{table}[tbp]
  \caption{The gauge charges of Higgses and fermions in the LR theory (Model A).}
  \begin{center}
    \begin{tabular}{|c|c|c|c|c|c|c|c|c|c|c|c|c|} \hline
                & $H_L$          & $H_R$         & $q_i$         & $\bar{q}_i$     & $\ell_i$        & $\bar{\ell}_i$ & $U_i$         & $\bar{U}_i$     & $D_i$          & $\bar{D}_i$     & $E_i$   & $\bar{E}_i$ \\ \hline
      $SU(3)_c$ & {\bf 1}        & {\bf 1}       & {\bf 3}       & ${\bf \bar{3}}$ & {\bf 1}         & {\bf 1}        & {\bf 3}       & ${\bf \bar{3}}$ & {\bf 3}        & ${\bf \bar{3}}$ & {\bf 1} & {\bf 1}     \\
      $SU(2)_L$ & {\bf 2}        & {\bf 1}       & {\bf 2}       & {\bf 1}         & {\bf 2}         & {\bf 1}        & {\bf 1}       & {\bf 1}         & {\bf 1}        & {\bf 1}         & {\bf 1} & {\bf 1}     \\
      $SU(2)_R$ & {\bf 1}        & {\bf 2}       & {\bf 1}       & {\bf 2}         & {\bf 1}         & {\bf 2}        & {\bf 1}       & {\bf 1}         & {\bf 1}        & {\bf 1}         & {\bf 1} & {\bf 1}     \\
      $U(1)_X$  & $\frac{1}{2}$ & $-\frac{1}{2}$ & $\frac{1}{6}$ & $-\frac{1}{6}$  & $- \frac{1}{2}$ & $\frac{1}{2}$  & $\frac{2}{3}$ & $-\frac{2}{3}$  & $-\frac{1}{3}$ & $\frac{1}{3}$   & $-1$    & $1$         \\ \hline
    \end{tabular}
  \end{center}
  \label{tab:chargesA}
\end{table}%

\begin{table}[tbp]
  \caption{The gauge charges of Higgses and fermions in the Mirror theory (Model D).}
  \begin{center}
    \begin{tabular}{|c|c|c|c|c|c|c|c|c|c|c|c|c|} \hline
                & $H_L$          & $H_R$         & $q_i$         & $q'_i$     & $\ell_i$        & $\ell'_i$ & $\bar{u}_i$         & $\bar{u}'_i$     & $\bar{d}_i$          & $\bar{d}'_i$     & $\bar{e}_i$   & $\bar{e}'_i$ \\ \hline
      $SU(3)_c$ & ${\bf 1}$        & {\bf 1}       & ${\bf 3}$       & ${\bf 3}$ & {\bf 1}         & {\bf 1}        & ${\bf \bar{3}}$       & ${\bf \bar{3}}$ & ${\bf \bar{3}}$        & ${\bf \bar{3}}$ & {\bf 1} & {\bf 1}     \\
      $SU(2)_L$ & {\bf 2}        & {\bf 1}       & {\bf 2}       & {\bf 1}         & {\bf 2}         & {\bf 1}        & {\bf 1}       & {\bf 1}         & {\bf 1}        & {\bf 1}         & {\bf 1} & {\bf 1}     \\
      $SU(2)_R$ & {\bf 1}        & {\bf 2}       & {\bf 1}       & {\bf 2}         & {\bf 1}         & {\bf 2}        & {\bf 1}       & {\bf 1}         & {\bf 1}        & {\bf 1}         & {\bf 1} & {\bf 1}     \\
      $U(1)_X$  & $\frac{1}{2}$ & $\frac{1}{2}$ & $\frac{1}{6}$ & $\frac{1}{6}$  & $- \frac{1}{2}$ & $-\frac{1}{2}$  & $-\frac{2}{3}$ & $-\frac{2}{3}$  & $\frac{1}{3}$ & $\frac{1}{3}$   & $1$    & $1$         \\ \hline
    \end{tabular}
  \end{center}
  \label{tab:chargesD}
\end{table}%

\subsection{Threshold correction}

Model-dependence arises from threshold corrections from the interactions that lead to the top quark Yukawa coupling. In Model A, these interactions are
\begin{align}
\label{eq:ytopA}
{\cal L} = - x \, q \bar{U}H_L  - x \, \bar{q} U H_R - M \bar{U} U.
\end{align}
After $SU(2)_R$ symmetry breaking, a linear combination of $\bar{u} \supset \bar{q}$ and $\bar{U}$ obtains a mass $\sqrt{x^2 v_R^2 + M^2}$ and the other linear combination remains massless to be identified with the right-handed top quark. The top Yukawa is
\begin{align}
    y_t = \frac{x^2 v_R}{\sqrt{x^2 v_R^2 + M^2}}.
\end{align}

The threshold correction to $\lambda_{\rm SM} (v_R)$ can be computed in the following way.
\begin{enumerate}
    \item Compute the Coleman-Weinberg potential of $H_L$ and $H_R$ from the top Yukawa sector and add it to the tree-level potential,
    \begin{align}
        \label{eq:generalPotential}
        V = \lambda (|H_L|^2 + |H_R|^2)^2 - m_H^2 (|H_L|^2 + |H_R|^2) + \epsilon |H_L|^2 |H_R|^2 + V_{\rm CW}(H_L, H_R).
    \end{align}
    \item
    Minimize the potential with respect to $H_R$ while taking $H_L=0$ to fix the parameter $m_H^2$ so that $\vev{H_R} = v_R$.
    \item
    Integrating out $H_R$ to obtain the effective potential $V_{\rm eff}(H_L)$ and fix the parameter $\epsilon$ to take the quadratic term of $H_L$ to be zero, which corresponds to electroweak fine-tuning.
    \item
    The coefficient of $|H_L|^4$ is of the form
    \begin{align}
    V_{\rm eff}/|H_L|^4 = a  - \frac{3}{16\pi^2} y_t^4 \, {\rm ln} \frac{|H_L^2|}{v_R^2},
    \end{align}
    where the constant term $a$ depends on the couplings of the UV theory.
    \item
    Comparing this with the CW potential of the SM in the $\overline{\rm MS}$ scheme,
    \begin{align}
        V_{\rm CW,SM}/|H_L|^4 = \lambda_{\rm SM}(\mu) -\frac{3}{16\pi^2} \, y_t^4 \left({\rm ln}\frac{y_t^2 |H_L|^2}{\mu^2} - \frac{3}{2}\right),
    \end{align}
    we obtain
    \begin{align}
        \lambda_{\rm SM}(v_R) = a +  \frac{3}{16\pi^2}y_t^4 \left({\rm ln}y_t^2  - \frac{3}{2}\right).
    \end{align}
\end{enumerate}
Applying this procedure to the Lagrangian in Eq.~\eqref{eq:ytopA}, we find
\begin{align}
    \lambda_{\rm SM}(v_R) = \frac{3}{16\pi^2} y_t^4 \left( - 2 + {\rm ln}y_t^2 + 2{\rm ln}\frac{x^2}{y_t^2} \right) \geq \frac{3}{16\pi^2} y_t^4 \left( - 2 + {\rm ln}y_t^2  \right), 
\end{align}
where the inequality is saturated when $x=y_t$, i.e., $M=0$. The prediction on $v_R$ for $M\neq0$ is always smaller than that for $M=0$.

In Model D, the top Yukawa is given by the Lagrangian
\begin{align}
\label{eq:ytopD}
{\cal L} = - x q \bar{u}H_L  - x q' \bar{u}' H_R - z q \bar{u}' H_L -  z q' \bar{u} H_R.
\end{align}
After Parity breaking, a linear combination of $\bar{u}$ and $\bar{u}'$ obtains a mass $\sqrt{x^2 + z^2} \, v_R$. Another linear combination remains massless and is identified as the right-handed top quark. The top Yukawa is
\begin{align}
    y_t= \frac{x^2 -z^2}{\sqrt{x^2 + z^2}}.
\end{align}
For $z\ll x$ and $x\ll z$, $y_t = x$ and $z$, respectively. 
We find that the threshold correction to $\lambda_{\rm SM}(v_R)$ is
\begin{align}
\lambda_{\rm SM}(v_R) = \frac{3}{16\pi^2} y_t^4 \left( - 2 + {\rm ln}y_t^2 + {\rm ln}\frac{x^2 + z^2}{y_t^2} \right) \geq \frac{3}{16\pi^2} y_t^4 \left( - 2 + {\rm ln}y_t^2  \right),
\end{align}
where the inequality is saturated when $z=0$ or $x=0$. The prediction on $v_R$ for $z\neq0$ or $x\neq0$ is always smaller than that for $z=0$ or $x=0$.

\subsection{Running of quartic}

In Model A where the right-handed SM fermions are mostly the Parity partners of the left-handed SM fermion, the RGE running of the Higgs quartic coupling is the same as the SM.
If we take the Dirac masses analogous to $M$ in Eq.~\ref{eq:ytopA} to be zero, the Parity partners of the SM fermions have masses $m_{\rm SM} v_R/v_L$ and the running of the $SU(3)_c\times U(1)_Y$ gauge couplings is modified.
In Model D, where the parity partners of the SM fermion are new particles, the Parity partners may be also as light as $m_{\rm SM} v_R/v_L$. 

\begin{figure}[tb]
    \centering
    \begin{minipage}{0.5\textwidth}
        \centering
        \includegraphics[width=.95\textwidth]{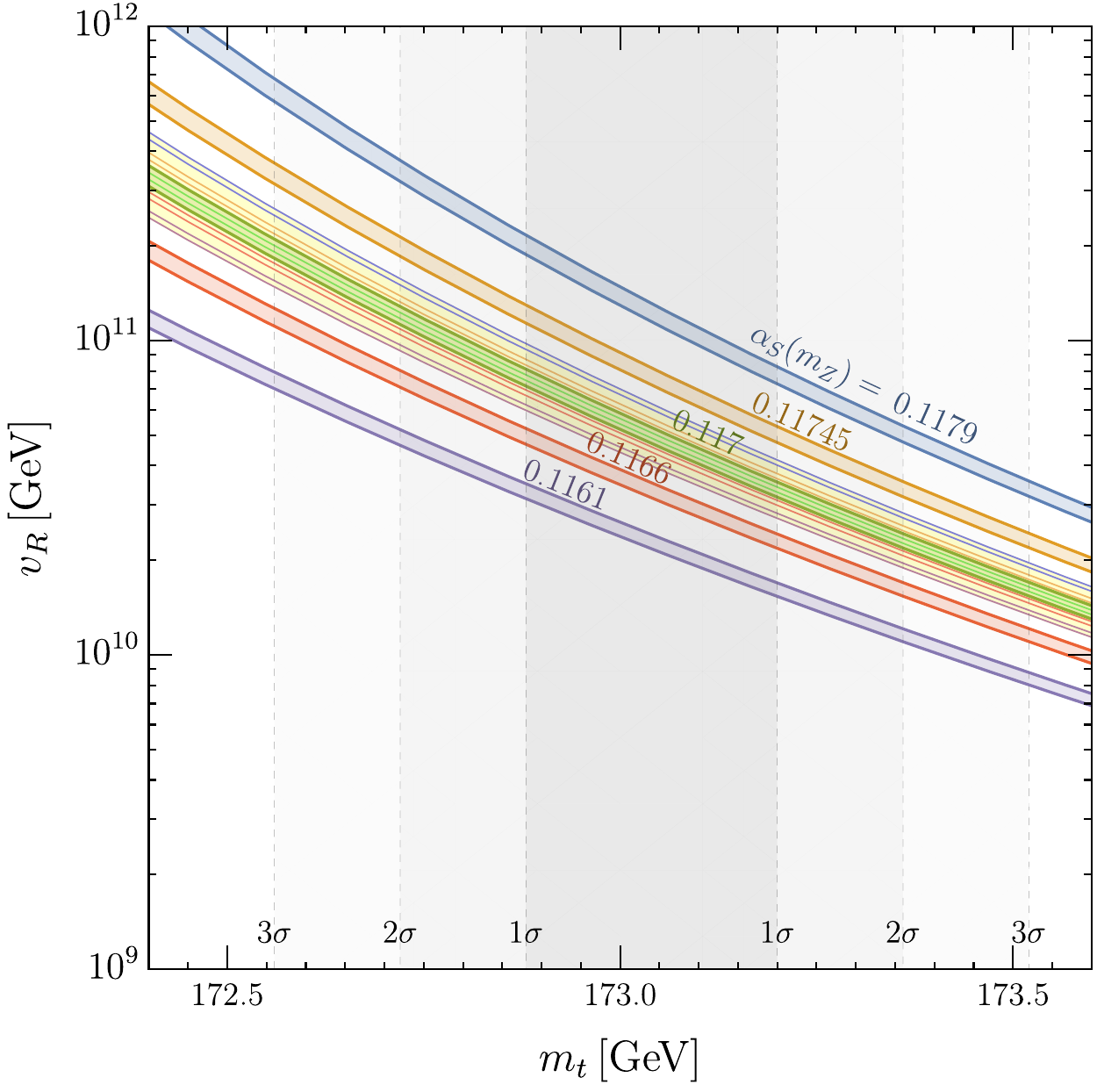} 
    \end{minipage}\hfill
    \begin{minipage}{0.5\textwidth}
        \centering
        \includegraphics[width=.95\textwidth]{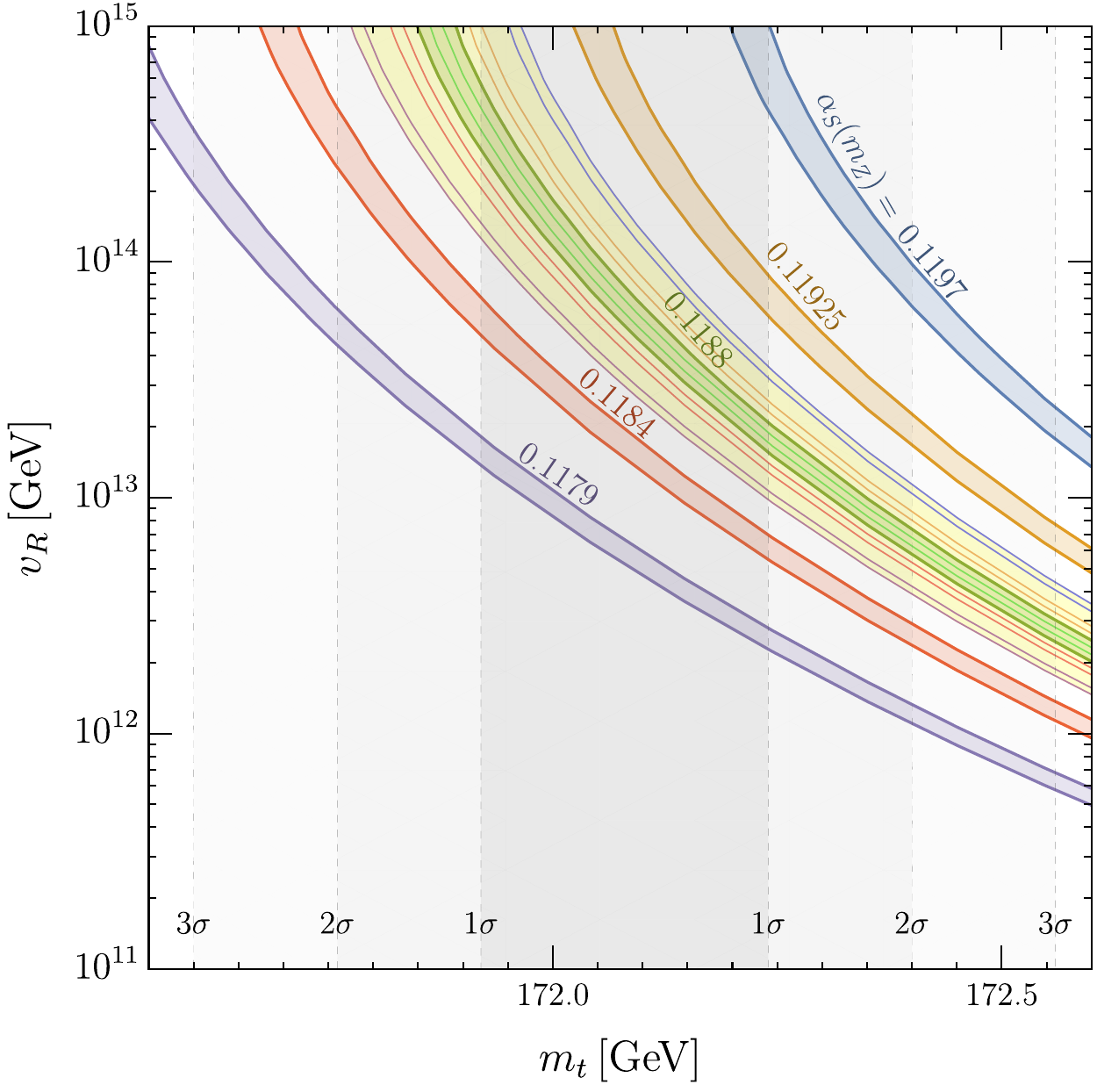} 
    \end{minipage}
    \caption{Future determinations of $v_R$ from improved measurements of SM parameters. The figures are analogous to Fig.~\ref{fig:vpPredictionFuture} but for the Mirror Theory. The two modifications are the the running of $\alpha_s$ from mirror quarks which are charged under the Standard Model $SU(3)$, as well as the modified threshold correction at $v_R$. \textbf{Left}: Centered on $m_t (\alpha_s)$ that are $1 \sigma$ above (below) the current values.   \textbf{Right}: Centered on $m_t (\alpha_s)$ that are $1 \sigma$ below (above) the current values. }
    \label{fig:vpPredictionFuturevp}
\end{figure}
Thus, the two main modifications to determining the running of the quartic in Model D are: 1) The incorporation of the mirror quarks to the running of $\alpha_s$ and 2) the incorporation of all mirror particles that possess hypercharge to the running of $\alpha_Y$. In both cases, we manually increment the number of light mirror fermions to the relevant beta functions at the scale $\mu = y_i' v_R$, where $y_i'$ are the mirror fermion Yukawa couplings. Note in the Mirror theory, the threshold correction Eq. \eqref{eq:thresholdCorrectionvR} to $\lambda$ at the scale $v_R$ is the same as in the Left-Right theory.
We numerically determine the renormalization scale at which $\lambda_{\rm SM}$ equals \eqref{eq:thresholdCorrectionvR} and show the results in Fig.~\ref{fig:vpPredictionFuturevp}, which is analogous to Fig.~\ref{fig:vpPredictionFuture} for the LR theory. In general, the incorporation of mirror quarks reduces the running of $\alpha_s$ at high scales, shifting the colored bands slightly up relative to Fig.~\ref{fig:vpPredictionFuture}.

\bibliographystyle{JHEP}
\bibliography{biblio}
\end{document}